\documentclass [aip, jmp, reprint, superscriptaddress, eqsecnum, onecolumn]{revtex4-1}
\usepackage {etex}

\usepackage [english]{babel}
\usepackage [T1]{fontenc}
\usepackage [utf8] {inputenc}
\usepackage {lmodern}
\usepackage {geometry}
\usepackage {amsfonts}
\usepackage {amsmath}	
\usepackage {amssymb}
\usepackage {amsxtra}
\usepackage [amsthm,thmmarks,amsmath]{ntheorem} 
\usepackage {wasysym}
\usepackage {mathtools}
\usepackage {calrsfs}
\usepackage {dsfont}
\usepackage {graphicx}

\usepackage {setspace}
\usepackage {color}
\usepackage [all,cmtip]{xy}
\usepackage {hyperref}
\usepackage [nottoc,numbib]{tocbibind}
\usepackage {diagbox}

\allowdisplaybreaks

\input diagxy

\DeclareFontFamily{OT1}{pzc}{}
\DeclareFontShape{OT1}{pzc}{m}{it}{<-> s * [1.30] pzcmi7t}{}
\DeclareMathAlphabet{\mathpzc}{OT1}{pzc}{m}{it}

\DeclareMathOperator{\sA}{\mathpzc A}
\DeclareMathOperator{\sB}{\mathpzc B}
\DeclareMathOperator{\sD}{\mathpzc D}

\DeclareMathOperator{\sK}{\mathpzc K}
\DeclareMathOperator{\sS}{\mathpzc S}

\DeclareMathOperator{\R}{\mathbb R}

\DeclareMathOperator{\C}{\mathbb C}
\DeclareMathOperator{\N}{\mathbb N}
\DeclareMathOperator{\Z}{\mathbb Z}

\DeclareMathOperator{\SW}{\textup SW}

\DeclareMathOperator{\fH}{\mathfrak H}

\DeclareMathOperator{\cB}{\mathcal B}

\DeclareMathOperator{\cS}{\mathcal S}

\DeclareMathOperator{\cO}{\mathcal O}

\DeclareMathOperator{\img}{img}

\DeclareMathOperator{\eff}{eff}
\DeclareMathOperator{\mol}{mol}
\DeclareMathOperator{\nuc}{nuc}

\DeclareMathOperator{\tr}{tr}

\newtheoremstyle{breakdef}%
  {\item[\rlap{\vbox{\normalfont\bfseries\hbox{\llap{##2}\hskip\labelsep
          ##1:}\hbox{\\[0.1cm]}}}]}%
  {\item[\rlap{\vbox{\normalfont\bfseries\hbox{\llap{##2}\hskip\labelsep
          ##1 (##3):}\hbox{\\[0.1cm]}}}]}
\newtheoremstyle{breaksatz}%
  {\item[\rlap{\vbox{\normalfont\normalsize\bfseries\hbox{\llap{##2}\hskip\labelsep
          ##1:}\hbox{\\[0.1cm]}}}]}%
  {\item[\rlap{\vbox{\normalfont\normalsize\bfseries\hbox{\llap{##2}\hskip\labelsep
          ##1 (##3):}\hbox{\\[0.1cm]}}}]}
\newtheoremstyle{breaklem}%
  {\item[\rlap{\vbox{\normalfont\normalsize\bfseries\hbox{\llap{##2}\hskip\labelsep
          ##1:}\hbox{\\[0.1cm]}}}]}%
  {\item[\rlap{\vbox{\normalfont\normalsize\bfseries\hbox{\llap{##2}\hskip\labelsep
          ##1 (##3):}\hbox{\\[0.1cm]}}}]}
\newtheoremstyle{breakprop}%
  {\item[\rlap{\vbox{\normalfont\normalsize\bfseries\hbox{\llap{##2}\hskip\labelsep
          ##1:}\hbox{\\[0.1cm]}}}]}%
  {\item[\rlap{\vbox{\normalfont\normalsize\bfseries\hbox{\llap{##2}\hskip\labelsep
          ##1 (##3):}\hbox{\\[0.1cm]}}}]}
\newtheoremstyle{breakbem}%
  {\item[\rlap{\vbox{\hbox{\hskip\labelsep\normalfont\bfseries
          ##1 ##2:}\hbox{\\[0.1cm]}}}]}%
  {\item[\rlap{\vbox{\hbox{\hskip\labelsep\normalfont\bfseries
          ##1 ##2 (##3):}\hbox{\\[0.1cm]}}}]}
\newtheoremstyle{breakbsp}%
  {\item[\rlap{\vbox{\hbox{\hskip\labelsep\normalfont\bfseries
          ##1 ##2:}\hbox{\\[0.2cm]}}}]}%
  {\item[\rlap{\vbox{\hbox{\hskip\labelsep\normalfont\bfseries
          ##1 ##2 (##3):}\hbox{\\[0.2cm]}}}]}
\newtheoremstyle{breakkor}%
  {\item[\rlap{\vbox{\hbox{\hskip\labelsep\normalfont\bfseries
          ##1 ##2:}\hbox{\\[0.1cm]}}}]}%
  {\item[\rlap{\vbox{\hbox{\hskip\labelsep\normalfont\bfseries
          ##1 ##2 (##3):}\hbox{\\[0.1cm]}}}]}
\newtheoremstyle{proof}%
  {\item[\rlap{\vbox{\hbox{\hskip\labelsep\normalfont\bfseries
          \underline{##1:}}\hbox{\\[0.1cm]}}}]}%
  {\item[\rlap{\vbox{\hbox{\hskip\labelsep\normalfont\bfseries
          \underline{##1 (##3):}}\hbox{\\[0.1cm]}}}]}
\theoremstyle{breakkor} 
\newtheorem{Definition}{Definition}[section] 
\theoremstyle{breakkor}

\theoremstyle{breakkor}

\theoremstyle{breakkor}

\theoremstyle{breakkor}

\theoremstyle{breakkor}
\theorembodyfont{\normalfont}
\newtheorem{Remark}[Definition]{Remark}

\theoremstyle{proof}
\theoremsymbol{\fbox{}}

\begin{document}
\title{Coherent states, quantum gravity and the Born-Oppenheimer approximation, I: General considerations}
\author{Alexander Stottmeister}
\email{alexander.stottmeister@gravity.fau.de}
\author{Thomas Thiemann}
\email{thomas.thiemann@gravity.fau.de}
\affiliation{Institut für Quantengravitation, Lehrstuhl für Theoretische Physik III, Friedrich-Alexander-Universität Erlangen-Nürnberg, Staudtstraße 7/B2, D-91058 Erlangen, Germany}
\begin{abstract}
This article, as the first of three, aims at establishing the (time-dependent) Born-Oppenheimer approximation, in the sense of space adiabatic perturbation theory, for quantum systems constructed by techniques of the loop quantum gravity framework, especially the canonical formulation of the latter. The analysis presented here fits into a rather general framework, and offers a solution to the problem of applying the usual Born-Oppenheimer ansatz for molecular (or structurally analogous) systems to more general quantum systems (e.g. spin-orbit models) by means of space adiabatic perturbation theory. The proposed solution is applied to a simple, finite dimensional model of interacting spin systems, which serves as a non-trivial, minimal model of the aforesaid problem. Furthermore, it is explained how the content of this article, and its companion, affect the possible extraction of quantum field theory on curved spacetime from loop quantum gravity (including matter fields).
\end{abstract}
\maketitle
\tableofcontents
\section{Introduction}
\label{sec:intro}
In this article, we begin our investigations into a framework that allows for the formulation of the (time-dependent) Born-Oppenheimer approximation for physical models of the type of loop quantum gravity \cite{ThiemannModernCanonicalQuantum, RovelliQuantumGravity}.\\
To this end, we continue and generalise certain ideas and proposals for the extraction of quantum field theory on curved spacetimes\cite{FullingAspectsOfQuantum, WaldQuantumFieldTheory, HollandsAxiomaticQuantumField, BrunettiTheGenerallyCovariant} from models of loop quantum gravity present in the literature\cite{SahlmannTowardsTheQFT1,SahlmannTowardsTheQFT2, GieselBornOppenheimerDecomposition}, and try to put them into a rigorous and (computationally) effective mathematical framework. The latter is provided by \textit{space adiabatic perturbation theory} as developed by Panati, Teufel and Spohn in \cite{PanatiSpaceAdiabaticPerturbation,TeufelAdiabaticPerturbationTheory} which is a mathematically precise formulation, by means of pseudo-differential calculus, of the intuitive content of the Born-Oppenheimer approximation \cite{BornZurQuantentheorieDer} in a time-dependent setting. Namely, the quantum system under consideration is split into slow and fast subsystems, and (partially) dequantised in the slow sector (deformation quantisation). As a result, an algebra of functions on a manifold (representing the slow subsystem) taking values in operators on the fast subsystem's Hilbert space is obtained. Moreover, the algebra of functions admits a non-commutative $\star$-product, which captures the operator product of the slow subsystem and has an expansion that exploits the separation of scales between the slow and fast sector. Then, assuming the spectral problem of the function, more precisely its principal part in an expansion w.r.t the separation of scales, quantising to the Hamiltonian of the system is under sufficient control, the time evolution operator is analysed by means of the $\star$-product and its expansion (see \cite{EmmrichGeometryOfThe} for a general method). This latter analysis corresponds to the perturbative analysis of molecular spectra by means of instantaneous electron configuration for fixed distributions of nuclei in the original Born-Oppenheimer setup. \\
For the time-independent Born-Oppenheimer approximation originally formulated in \cite{BornZurQuantentheorieDer}, as opposed to the time-dependent setting, the use of pseudo-differential operators to construct locally isospectral effective Hamiltonians has some tradition. Classical works in this respect are \cite{KleinOnTheBornOppenheimer, CombesTheBornOppenheimer, HagedornHigherOrderCorrectionsI, HagedornHigherOrderCorrectionsIIa, HagedornHigherOrderCorrectionsIIb}. A general reduction scheme in the time-independent case, which is mathematically similar to space adiabatic perturbation theory, and therefore works for the more general type of models, we have in mind, as well, is depicted in \cite{MartinezOnAGeneral}.\\[0.1cm]
Now, let us describe, in a little more detail, in which sense space adiabatic perturbation theory provides a suitable framework for the (time-dependent) Born-Oppenheimer approximation in models {\`a} la loop quantum gravity, and how quantum field theory on curved spacetimes fits into this perspective:\\[0.1cm]
The Born-Oppenheimer has a long tradition in applications to quantum gravity\footnote{Cf. \cite{KieferQuantumGravity} for a historical overview.}. But, its use in the context of loop quantum gravity is only quite recent, accompanying the construction of (effective) models. A first attempt to incorporate the Born-Oppenheimer approximation into the canonical formulation of loop quantum gravity was made in \cite{GieselBornOppenheimerDecomposition} (an application in covariant loop quantum gravity/spin foam models can be found in \cite{RovelliSteppingOutOf}).\\
It is suggested in \cite{GieselBornOppenheimerDecomposition}, that the connection between loop quantum gravity models with matter content and quantum field theory on curved spacetimes, roughly, arises in the following way: Firstly, a deparametrised Hamiltonian formulation of the considered model in the presence of additional, so-called \textit{dust fields}, which provide a (physical) system of spacetime coordinates\cite{BrownDustAsA}\footnote{See \cite{GieselManifestlyGaugeInvariant1, GieselManifestlyGaugeInvariant2, GieselScalarMaterialReference} for the use in loop quantum gravity}, is chosen and quantised by the methods of loop quantum gravity. The deparametrisation is a generalisation of a similar procedure used to obtain the \textit{Friedmann equations} in cosmological models, where the (approximately) homogeneous and isotropic distribution of (super-)galactic structures in the universe serves as a dust field. Secondly, the quantum system is separated into slow and fast subsystems in accordance with the splitting into a gravitational and matter sector (in general, a mixing of gravitational and matter degrees of freedom is conceivable). Such a separation is motivated, on the one hand, by the fact that the natural mass scales, set by the coupling constants $\kappa$ (Einstein's constant) and $\lambda$ (scalar field), are typically well-separated, which is captured by the (small) dimensionless parameter $\varepsilon^{2}=\frac{\kappa}{\lambda}$, and, on the other hand, by the observation that all experiments on gravity-matter systems, performed so far, are well described by treating the gravitational field as a classical entity (notably, the accordance  of (classical) $\Lambda$-CDM model with recent observations of the cosmic microwave background by PLANCK). In specific models, a further investigation of the separation of scales is necessary, as it has to be ensured that the physical states subject to analysis respect this formal argument\footnote{Thinking of the original Born-Oppenheimer ansatz applied to molecular Hamiltonians, it is easily understood that the separation of scales between the electrons and the nuclei quantified by the mass ratio $\varepsilon^{2}=\frac{m_{e}}{m_{\nuc}}$ is not sufficient to justify the usual slow-fast decomposition, but additional bounds on the total energy, and thus on the kinetic energy, of the molecular system are necessary to conclude that time scale for the motion of the nuclei is, indeed, much larger than that for the electrons\cite{TeufelAdiabaticPerturbationTheory}.}. Thirdly, the Born-Oppenheimer ansatz is invoked to obtain effective Hamiltonians for the (fast) matter sector, which are parametrised by (classical) configurations (or states) of the slow system. Ideally, these effective Hamiltonians define quantum field theories for given (external) classical gravitational fields, i.e. quantum field theories on curved spacetimes. Finally, information on the spectral problem of the effective Hamiltonians is used as input for the description of the total quantum system.\\[0.1cm]
As pointed out in \cite{GieselBornOppenheimerDecomposition}, two main obstacles to a successful implementation of the outlined program present themselves in the following form:
\begin{itemize}
	\item[1.] \textbf{Non-commutative fast-slow coupling:}\\[0.1cm]
The third step, i.e. applying the Born-Oppenheimer ansatz in the construction of effective Hamiltonians, requires a peculiar structure of the Hamiltonian of the coupled quantum system: The part of the Hamiltonian modelling the coupling between the slow and fast subsystems needs to implemented by a family of mutually commuting, self-adjoint operators w.r.t. the slow variables. This property implies that the coupling-part of the Hamiltonian admits a description as a fibred operator over some parameter space connected with the slow variables, i.e. the common spectrum of the slow-sector operators mediating the coupling.\\
Due to the structure of the quantum algebra, the \textit{holonomy-flux algebra} (or its spin-offs), constructed in the quantisation of the gravitational field along the lines of loop quantum gravity, the Hamiltonians of the models, we are interested, do not have this feature: The operators representing the (spatial) metric, the \textit{flux operators} (short: fluxes), generate a non-commutative (sub)algebra, and it is the (spatial) metric that couples to the matter fields in generic gravity-matter Hamiltonians.
	\item[2.] \textbf{Continuum limit:}\\[0.1cm]
Quantisations {\`a} la loop quantum gravity of classical field theories including gravity are modelled on a projective limit, $\overline{\Gamma}=\varprojlim_{i\in I}\Gamma_{i}$, of truncated configuration or phase spaces, $\Gamma_{i},\ i\in I$. Typically, $\overline{\Gamma}$ has an interpretation as a distributional completion of the classical smooth configuration or phase space, $\Gamma$.
Thus, in the quantum theory only $\overline{\Gamma}$ is naturally accessible, and the specification of elements of $\Gamma$, which contains the classical gravitational field configurations or states, has to be achieved via observables admitting a suitable continuum limit.\\
It is, therefore, a minimal requirement that we find a generalisation of the Born-Oppenheimer ansatz, which is compatible with the projective limit structure arising in the quantisation process.
\end{itemize}
In the present article, we focus on a possible resolution of the first issue on rather general grounds by means of space adiabatic perturbation theory, which offers a more flexible framework than the original Born-Oppenheimer ansatz. The second problem will be (partly) addressed in our second and third article\cite{StottmeisterCoherentStatesQuantumII, StottmeisterCoherentStatesQuantumIII}, where we also establish the mathematical basis necessary to realise the program of space adiabatic perturbation theory in models, which are structurally similar to loop quantum gravity.
\\[0.25cm]
The remainder of the article is structured as follows:\\[0.1cm]
In section \ref{sec:boa}, we recall, in an informal way, the Born-Oppenheimer ansatz (e.g. \cite{ChruscinskiGeometricPhasesIn}), as it usually presented for molecular Hamiltonians (or systems analogous to those), and the derivation of effective Hamiltonians, which govern the motion of the fast subsystem inside the adiabatically decoupled subspaces and allow for the derivation of effective equations for the slow variables in semi-classical limit. Following this, we argue that the treatment of more general quantum systems, which allow for a splitting into slow and fast degrees of freedom, but do not have the peculiar form common to molecular Hamiltonians, requires a generalisation of the Born-Oppenheimer ansatz. This generalisation manifests itself in extending the (de-)quantisations procedure in terms of orthogonal (in the generalised sense) pure state families (\textit{fibered} or \textit{direct integral representations of operators}), which is at the heart of the original Born-Oppenheimer ansatz, to more general deformation (de-)quantisations, e.g. coherent pure state quantisations (\textit{Wick/Anti-Wick} or \textit{Berezin quantisations}) or \textit{Kohn-Nirenberg} and \textit{Weyl quantisations}.\\
In section \ref{sec:weyl}, we formulate the framework of space adiabatic perturbation in rather general terms, rather focusing on structural aspects than on technical details, due to the fact that its original formulation \cite{PanatiSpaceAdiabaticPerturbation} is given in the context of Weyl quantisation on $\R^{2d}$ for operators on $L^{2}(\R^{d})$, which is a setting to narrow for the applications that we have in mind.\\
In section \ref{sec:spin}, we apply the general framework of the previous section to a simple, finite-dimensional model of two coupled spin systems introduced by Faure and Zhilinskii \cite{FaureTopologicalPropertiesOf} in a discussion of topological aspects of the Born-Oppenheimer approximation. The (de-)quantisation we choose to analyse this model of coupled spin systems is the so-called \textit{Stratonovich-Weyl quantisation} for the 2-sphere $S^{2}$\cite{StratonovichOnDistributionsIn, VarillyTheMoyalRepresentation}, which is a direct analogue of the Weyl quantisation on $\R^{2}$. The reason, why we discuss this model, is that it constitutes a sort of minimal representative of a quantum system, that is not amenable to the usual Born-Oppenheimer approximation because of the structure of the $\mathfrak{su}_{2}$-algebras describing its observables. Furthermore, the non-trivial topology of the manifold $S^{2}$ affects the applicability of space adiabatic perturbation theory in an interesting way\footnote{Cf. \cite{FreundEffectiveHamiltoniansFor} for a discussion in the context of periodic Schrödinger operators with external field and magnetic Bloch bands} -- a feature that is expected for loop quantum gravity models, which are based on projective limits of co-tangent bundles $T^{*}G$ of compact Lie groups $G$, as well. Due to compactness of $S^{2}$ , reflecting the finite dimensionality of the model, it is not of vital importance to pay to much attention to the technical details of the (de-)quantisation procedure (all operators are bounded, $C^{\infty}(S^{2})=C^{\infty}_{b}(S^{2})$).\\
Finally, we conclude the article in section \ref{sec:con}, and comment on the implications of our findings, especially in respect of our companion articles \cite{StottmeisterCoherentStatesQuantumII, StottmeisterCoherentStatesQuantumIII}.
\section{On the Born-Oppenheimer ansatz}
\label{sec:boa}
In this section, we discuss aspects of the time-dependent Born-Oppenheimer approximation in the analysis of coupled quantum systems,
\begin{align}
\label{eq:slowfastsector}
\fH & = \fH_{s}\otimes\fH_{f},
\end{align}
consisting of two sectors characterised by well-separated interaction/time scales (captured by a ``small'' parameter $\varepsilon$), hereafter called the \textit{slow sector} or \textit{slow degrees of freedom}, $\fH_{s}$, and the \textit{fast sector} or \textit{fast degrees of freedom}, $\fH_{f}$. We explain on a rather formal level in which sense the conventional Born-Oppenheimer approximation\cite{ChruscinskiGeometricPhasesIn} fits into the picture of \textit{pure state (de-)quantisation}\cite{SimonTheClassicalLimit, LandsmanMathematicalTopicsBetween} w.r.t. to a (generalised) orthogonal family of pure states in $\fH_{s}$ (or a suitable extension $\fH_{f}\subset\mathfrak{S}^{'}_{f}$), which is adapted to the operators of the slow sector that couple non-trivially to the fast sector. As observed in \cite{GieselBornOppenheimerDecomposition}, it turns out, that it is of vital importance, that the coupling operators are assumed to be mutually commuting, for this approach to work.\\
Following this, we argue that a treatment of coupled quantum systems, where this restrictive assumption is not satisfied (e.g. the Dirac equation with slowly varying external fields \cite{PanatiSpaceAdiabaticPerturbation}, spin-orbit coupling \cite{FaureTopologicalPropertiesOf}), requires another type of (de-)quantisation, presumably not even by pure states.
\subsection{The Born-Oppenheimer ansatz}
\label{subsec:oboa}
The ansatz of Born and Oppenheimer \cite{BornZurQuantentheorieDer} is usually derived in the context of molecular Hamiltonians with external magnetic field \cite{ChruscinskiGeometricPhasesIn, PanatiTheTimeDependent},
\begin{align}
\label{eq:molecularHamiltonian}
\hat{H}_{\mol} & = \frac{1}{2 m_{\nuc}}\left(\hat{P}+A(\hat{Q})\right)^{2} + H_{e}(\hat{Q};\hat{q},\hat{p}),
\end{align}
defined on a dense domain $D(H_{\mol})\subset L^{2}(\R^{d},\fH_{f})\cong L^{2}(\R^{d})\otimes\fH_{f}$ and self-adjoint there. The slow nuclei are modelled on $\{(Q,P),L^{2}(\R^{d})\}$ and the fast electrons constitute the fibre Hilbert space $\{(q,p),\fH_{f}\}$. \\
For simplicity, we do not include any (possibly internal) degrees of freedom besides position and momentum of the nuclei and electrons into the discussion. \\
Due to the rather special form of \eqref{eq:molecularHamiltonian}, i.e. the coupling between slow and fast degrees of freedom happens solely via the vector of mutually commuting operators $\hat{Q}$, and $H_{e}(\hat{Q};\hat{q},\hat{p})$ is fibered over the spectrum $\sigma(Q)=\R^{d}$ of $Q$, it is possible to analyse the spectral properties of $\hat{H}_{\mol}$ by means of spectral decomposition of the (self-adjoint) electronic Hamiltonians $H(Q;\hat{q},\hat{p}),\ Q\in\R^{n}$ for fixed configurations of the nuclei. Namely, we introduce the a $Q$-dependent orthonormal basis of $\fH_{f}$ (or at least of a subspace of bound states $\fH^{b}_{f}$):
\begin{align}
\label{eq:molecularfibrebasis}
H(Q;\hat{q},\hat{p})\psi_{n, d_{n}}(Q) & = e_{n}(Q)\psi_{n, d_{n}}(Q) ,\ \ \ \psi_{n, d_{n}}(Q) \in\fH_{f},\ Q\in\R^{d},\ n\in N,d_{n}\in D_{N},
\end{align}
where a discrete, possibly degenerate, fibered spectrum, $\{\{e_{n}(Q)\}_{n\in N}\}_{Q\in\R^{d}}$, without eigenvalue crossings is assumed to exists for the family $\{H(Q;\hat{q},\hat{p})\}_{Q\in\R^{d}}$. $\{e_{n}(Q)\}_{Q\in\R^{d}}$ is called the \textit{n-th electronic band}. Next, we introduce the (generalised) complete product basis
\begin{align}
\label{eq:boabasis}
\{\delta^{(d)}_{Q}\otimes\psi_{n, d_{n}}(Q)\}_{Q\in\R^{d}}\subset L^{2}(\R^{d},\fH_{f}),
\end{align}
and project to the component equations of the eigenvalue equation,
\begin{align}
\label{eq:moleculareigenvalueequation}
\hat{H}_{\mol}\Psi^{E} & = E\Psi^{E},\ \ \ \Psi^{E}\in L^{2}(\R^{d},\fH_{f}),
\end{align}
w.r.t. this basis ($\hbar=1$):
\begin{align}
\label{eq:moleculareigenvalueprojection}
E\ \Psi^{E}_{n,d_{n}}(Q) & = E \left(\delta^{(d)}_{Q}\otimes\psi_{n, d_{n}}(Q),\Psi^{E}\right) = \left(\delta^{(d)}_{Q}\otimes\psi_{n, d_{n}}(Q),\hat{H}_{\mol}\Psi^{E}\right) \\ \nonumber
 & = \frac{1}{2 m_{\nuc}}\left(\delta^{(d)}_{Q}\otimes\psi_{n, d_{n}}(Q),\left(\hat{P}+A(\hat{Q})\right)^{2}\Psi^{E}\right)  + \left(\delta^{(d)}_{Q}\otimes\psi_{n, d_{n}}(Q),\hat{H}_{e}(\hat{Q};\hat{q},\hat{p})\Psi^{E}\right) \\ \nonumber
 & = \sum_{n'',d_{n''}}\left(-\frac{1}{2 m_{\nuc}}\sum_{n',d_{n'}}\sD^{n',d_{n'}}_{n,d_{n}}\cdot\sD^{n'',d_{n''}}_{n',d_{n'}}+ e_{n}(Q)\delta_{n,n''}\delta_{d_{n},d_{n''}}\right)\Psi^{E}_{n'',d_{n''}}(Q),
\end{align}
where we used the resolution of unity
\begin{align}
\label{eq:molecularresolutionofunity}
\mathds{1}_{L^{2}(\R^{d},\fH_{f})} & = \int_{\R^{d}}dQ\sum_{n,d_{n}}(\delta^{(d)}_{Q}\otimes\psi_{n,d_{n}}(Q))\otimes\left(\delta^{(d)}_{Q}\otimes\psi_{n,d_{n}}(Q),\ .\ \right),
\end{align}
and defined
\begin{align}
\label{eq:molecularcovder}
\sD^{n',d_{n'}}_{n,d_{n}} & := \delta_{n,n'}\delta_{d_{n},d_{n'}}(\nabla_{Q}+iA(Q))-i\sA(Q)^{n',d_{n'}}_{n,d_{n}}, \\
\label{eq:molecularpregaugefield}
\sA(Q)^{n',d_{n'}}_{n,d_{n}} & := i\left(\psi_{n,d_{n}}(Q),(\nabla_{Q}\psi_{n',d_{n'}})(Q)\right)_{\fH_{f}}.
\end{align}
The adiabatic Born-Oppenheimer approximation of \eqref{eq:moleculareigenvalueprojection}, improved by the \textit{Berry-Simon connection} \cite{BerryQuantalPhaseFactors, SimonHolonomyTheQuantum} or \textit{Mead potential} \cite{MeadOnTheDetermination}, consists in ignoring the inter-band terms of the operator $\sD$, which yields (in electronic units $m_{e}=1,\ \tfrac{m_{e}}{m_{\nuc}}=\varepsilon^{2}$) 
\begin{align}
\label{eq:originalBOA}
E\ \Psi^{E}_{n,d_{n}}(Q) & = \sum_{d''_{n}}\Bigg(-\frac{\varepsilon^{2}}{2}\sum_{d'_{n}}(\sD_{n})^{d'_{n}}_{d_{n}}\cdot(\sD_{n})^{d''_{n}}_{d'_{n}}+ e_{n}(Q)\delta_{d_{n},d''_{n}}\Bigg)\Psi^{E}_{n,d''_{n}}(Q),
\end{align}
where $\sD_{n}$ is the diagonal part of $\sD$ in n-th electronic band. Additionally, a perturbative expansion in the \textit{adiabatic parameter} $\varepsilon$ of the eigenvalues $E$ and eigenvector coefficients $\Psi^{E}_{n,d_{n}}(Q)$ is performed\footnote{In their original paper Born and Oppenheimer perform the perturbative expansion in $\sqrt{\varepsilon}$, but they only obtain contribution in zeroth, second and fourth order \cite{BornZurQuantentheorieDer}. Later, mathematical rigorous treatments of the perturbative expansion make use of $\varepsilon$ \cite{KleinOnTheBornOppenheimer}.}.\\[0.1cm]
A slightly more educated guess, which is in spirit of Kato's time-adiabatic theorem, proceeds by projecting out the intra-band parts of the operator on the right hand side of \eqref{eq:moleculareigenvalueprojection}. This amounts to using the band projections,
\begin{align}
\label{eq:molecularbandprojection}
\hat{\Pi}_{n} & = \int_{\R^{d}}dQ\ \Pi_{n}(Q)\ \delta^{(d)}_{Q}\otimes\left(\delta^{(d)}_{Q},\ .\ \right), & \Pi_{n}(Q) = \sum_{d_{n}}\psi_{n,d_{n}}(Q)\otimes\left(\psi_{n,d_{n}}(Q),\ .\ \right)_{\fH_{f}},
\end{align}
which commute with $H_{e}(\hat{Q};\hat{q},\hat{p})$, to define the intra-band effective Hamiltonians:
\begin{align}
\label{eq:moleculareffectiveHamiltonian}
\hat{H}^{\eff}_{\mol,n} & = \hat{\Pi}_{n}\hat{H}_{\mol}\hat{\Pi}_{n} \\ \nonumber
 & = \int_{\R^{d}}dQ\sum_{d_{n},d'_{n}}\left(\delta^{(d)}_{Q}\otimes\psi_{n,d_{n}}\right)\left(\hat{\Pi}_{n}\hat{H}_{\mol}\hat{\Pi}_{n}\right)^{d'_{n}}_{d_{n}}(Q,-i\varepsilon\nabla_{Q})\left(\delta^{(d)}_{Q}\otimes\psi_{n,d'_{n}},\ .\ \right),
\end{align}
where
\begin{align}
\left(\hat{\Pi}_{n}\hat{H}_{\mol}\hat{\Pi}_{n}\right)^{d'_{n}}_{d_{n}}(Q,-i\varepsilon\nabla_{Q}) & = \Bigg(\left(-i\varepsilon\nabla_{Q}+\varepsilon A(Q)-\varepsilon\sA(Q)_{n}\right)^{2} \\[0.25cm] \nonumber
 & \hspace{0.5cm}+\frac{\varepsilon^{2}}{2}\sum_{n'\neq n}\left(\sA(Q)^{n'}_{n}\cdot\sA(Q)^{n}_{n'}\right)+e_{n}(Q)\Bigg)^{d'_{n}}_{d_{n}}.
\end{align}
$\sA(Q)_{n}$ is the quantum-geometric magnetic potential and $\Phi(Q)_{n} = \sum_{n'\neq n}\sA(Q)^{n'}_{n}\cdot\sA(Q)^{n}_{n'}$ is the quantum-geometric electric potential. The relevance of the band projections and the effective Hamiltonians lies within the fact, that the band subspaces are invariant to first order in $\varepsilon$ for states of bounded kinetic energy, i.e.
\begin{align}
\label{eq:firstorderinvariantsubspace}
[\hat{H}_{\mol},\hat{\Pi}_{n}] & = \cO(\varepsilon).
\end{align}
Although, there is a subtlety associated with \eqref{eq:firstorderinvariantsubspace}, because the relevant time scale for the slow sector is of order $\varepsilon^{-1}$ (\textit{Duhamel's formula} and \textit{gap conditions} are important here\cite{PanatiTheTimeDependent}).  \\[0.1cm]
First order time-adiabatic theorems concerning the approximation of the intra-band dynamics by means of the first order expansion of \eqref{eq:moleculareffectiveHamiltonian} in $\varepsilon$ (or its generalisations to collections of electronic bands) can be found in e.g. \cite{SpohnAdiabaticDecouplingAnd, TeufelAdiabaticPerturbationTheory, PanatiTheTimeDependent}. A systematic treatment of higher order corrections requires more refined techniques to be discussed in section \ref{sec:weyl}. Especially, the naive hope that the second order expansion of \eqref{eq:moleculareffectiveHamiltonian} in $\varepsilon$ is compatible with a second order adiabatic theorem is not justified\cite{TeufelAdiabaticPerturbationTheory, PanatiTheTimeDependent}. Fortunately, these techniques also lift the restriction of the above considerations to couplings of slow and fast degrees of freedom via mutually commuting operators in the slow sector.\\[0.1cm]
Semi-classical approximations to the dynamics of the slow variables can be obtained by $\varepsilon$-dependent pseudo-differential techniques (\textit{Egorov's theorem}\cite{RobertAutourDL, PanatiSpaceAdiabaticPerturbation, TeufelAdiabaticPerturbationTheory}), yielding in zeroth order in $\varepsilon$ classical dynamics governed by a Hamiltonian with potential energy given by the electronic energy associated with the band (\textit{Peierls substitution}), i.e.
\begin{align}
\label{eq:peierlsHamiltonian}
H_{\mol,0}(P,Q) & = \frac{1}{2}P^{2} + e_{n}(Q).
\end{align}
In the case of coupled observables on the slow and fast sector, we have to take into account first oder corrections to \eqref{eq:peierlsHamiltonian} (matrix-valued in the presence of degeneracies):
\begin{align}
\label{eq:peierlsHamiltonianfirstorder}
H_{\mol,(1)}(P,Q) & = \frac{1}{2}P^{2} + 2\varepsilon P(A(Q)-\sA(Q)_{n}) + e_{n}(Q).
\end{align}
\subsection{The coherent-state Born-Oppenheimer ansatz}
\label{subsec:csboa}
As pointed out in the previous section, the conventional Born-Oppenheimer approximation exploits to some extent the special structure of \eqref{eq:molecularHamiltonian} as being a perturbation, which acts on the slow degrees of freedom, of an operator, $H_{e}(\hat{Q};\hat{q},\hat{p})$, fibered over the common spectrum of the mutually commuting operators $\hat{Q}_{i},\ i=1,...,d,$ of the slow sector, which provide the coupling to the fast degrees of freedom.\\[0.1cm]
But, for the applications, that we have in mind, it would be advantageous to lift the restriction on the mutual commutativity of the coupling operators $Q_{i},\ i=1,...,d$. Clearly, if we do not require the vector $Q$ to have mutually commuting components, we will have to find a viable substitute for the (generalised) product basis \eqref{eq:boabasis}. A property that makes \eqref{eq:boabasis} special, is that it provides a diagonalisation of the slow-fast coupling operator $H_{e}(\hat{Q};\hat{q},\hat{p})$, which supports the idea that electronic configuration $\psi_{n,d_{n}}(Q)$ for fixed configuration of nuclei $Q\in\R^{d}$ can be used to determine the properties of a molecule.\\
One possibility to generalise this aspect to encompass the non-commutative setting is to move from a configuration space approach to the slow variables to a phase space approach. More precisely, if we treat $H_{e}(\ .\,\hat{q},\hat{p}):\R^{d}\rightarrow L(\fH_{f})$ as a function on configuration space $\R^{d}$ subjected to the (operator-valued) pure state quantisation\cite{LandsmanMathematicalTopicsBetween},
\begin{align}
\label{eq:configurationpurestatequant}
F(\R^{d},L(\fH_{f})) & \rightarrow L(L^{2}(\R^{d},\fH_{f})), & f(\ .\,\hat{q},\hat{p}) & \mapsto \int_{\R}dQ\ f(Q,\hat{q},\hat{p})\delta^{(d)}_{Q}\otimes\left(\delta^{(d)}_{Q},\ .\ \right),
\end{align}
w.r.t. the system of (generalised orthonormal) pure states $\{\delta^{(d)}_{Q}\}_{Q\in\R^{d}}$, we replace the system of pure states or even the type of quantisation to handle (operator-valued) functions $F(T^{*}\R^{d},L(\fH_{f}))$ on phase space $T^{*}\R^{d}\cong\R^{2d}$. The reason for this lies within the fact that phase space quantisation typically allows for a wider range of operators to be covered, as the orthogonality relation at the heart of \eqref{eq:configurationpurestatequant} are lifted\cite{SimonTheClassicalLimit}. Another important point to make in this respect is, that we are not so much interested in a quantisation scheme, but a de-quantisation scheme, i.e. a way to assign functions (also called \textit{symbols}), $f\in F(T^{*}\R^{d},L(\fH_{f}))$, on the phase space of the slow degrees of freedom with values in operators on the Hilbert space of the fast sector.\\
The goal is then, assuming the spectral problem of the operators $\{f(Q,P;\hat{q},\hat{p})\}_{(Q,P)\in\R^{2d}}$ is under sufficient control, to systematically approximate the full spectral problem and the dynamics of $f(\hat{Q},\hat{P};\hat{q},\hat{p})$, in terms of configurations of the fast degrees of freedom for fixed phases or instantaneous states of the (classical) slow sector:
\begin{align}
\label{eq:instantaneousfibrebasis}
f(Q,P;\hat{q},\hat{p})\psi_{n,d_{n}}(Q,P) & = e_{n}(Q,P)\psi_{n,d_{n}}(Q,P)
\end{align}
for $\psi_{n, d_{n}}(Q,P) \in\fH_{f}, (Q,P)\in\R^{2d}, n\in N,d_{n}\in D_{N},$ in analogy with \eqref{eq:molecularfibrebasis}.\\[0.1cm]
Before, we turn to the discussion of a systematic treatment of the outlined ideas in the framework of (space) adiabatic perturbation theory \cite{PanatiSpaceAdiabaticPerturbation, TeufelAdiabaticPerturbationTheory}, we take a look at what will happen to \eqref{eq:moleculareigenvalueprojection}, if we replace \eqref{eq:configurationpurestatequant} by a \textit{coherent state quantisation}\cite{LandsmanMathematicalTopicsBetween, BerezinQuantization, KlauderCoherentStatesApplications, GazeauCoherentStatesIn} for electronic energies depending on the momenta of the nuclei as well, $H_{e}(\hat{Q},\hat{P};\hat{q},\hat{p})$.\\
To this end, we assume that $H_{e}(\hat{Q},\hat{P};\hat{q},\hat{p})$ admits an \textit{upper or contravariant symbol}, $H_{e,\varepsilon}(\ .\ ;\hat{q},\hat{p})$, w.r.t. the ($\varepsilon$-dependent) standard coherent states $\zeta^{(\varepsilon)}_{Z}\in L^{2}(\R^{d}),\ Z\in\C^{d},$ for the CCR algebra $\{\hat{A}_{\varepsilon} = \tfrac{1}{\sqrt{2}}(\hat{Q}+i\hat{P}), \hat{A}^{*}_{\varepsilon} = \tfrac{1}{\sqrt{2}}(\hat{Q}-i\hat{P}), [\hat{A}_{\varepsilon},\hat{A}^{*}_{\varepsilon}] = \varepsilon\}$ associated with the $\varepsilon$-scaled slow sector $\{\hat{Q}_{\varepsilon},\hat{P}_{\varepsilon},[\hat{P},\hat{Q}]=-i\varepsilon\}$\footnote{In the previous subsection, we worked in the representation $\hat{Q}_{\varepsilon}=Q\cdot,\ \hat{P}_{\varepsilon}=-i\varepsilon\nabla_{Q}$.}:
\begin{align}
\label{eq:electronicuppersymbol}
H_{e}(\hat{Q}_{\varepsilon},\hat{P}_{\varepsilon};\hat{q},\hat{p}) = H_{e}(\hat{A}_{\varepsilon},\hat{A}^{*}_{\varepsilon};\hat{q},\hat{p}) & = \int_{\C^{d}}\frac{d^{2}Z}{(\varepsilon\pi)^{d}}\ H_{e,\varepsilon}(Z,\overline{Z};\hat{q},\hat{p})\ \zeta^{(\varepsilon)}_{Z}\otimes\left(\zeta^{(\varepsilon)}_{Z},\ .\ \right)_{L^{2}(\R^{d})}.
\end{align}
Next, we assume the existence of a family of bases $\{\psi^{(\varepsilon)}_{n,d_{n}}(Z,\overline{Z})\}_{Z\in\C^{d}}$ of $\fH_{f}$ (or a subspace thereof) adapted to the symbol $H_{e,\varepsilon}(\ .\ ;\hat{q},\hat{p})$ in the sense of \eqref{eq:instantaneousfibrebasis}, from which construct the (overcomplete) family of product states $\{\zeta^{(\varepsilon)}_{Z}\otimes\psi^{(\varepsilon)}_{n,d_{n}}(Z,\overline{Z})\}_{Z\in\C^{d}}\subset L^{2}(\R^{d},\fH_{f})$. Stressing the analogy with the previous subsection further, we introduce the (self-adjoint) complete family of operators:
\begin{align}
\label{eq:coherentbandprojections}
\hat{\Pi}^{\varepsilon}_{n} & = \int_{\C^{d}}\frac{d^{2}Z}{(\varepsilon\pi)^{d}}\Pi^{(\varepsilon)}_{n}(Z,\overline{Z})\ \zeta^{(\varepsilon)}_{Z}\otimes\left(\zeta^{(\varepsilon)}_{Z},\ .\ \right)_{L^{2}(\R^{d})}, \\ \nonumber
\Pi^{(\varepsilon)}_{n}(Z,\overline{Z}) & = \sum_{d_{n}}\psi^{(\varepsilon)}_{n,d_{n}}(Z,\overline{Z})\otimes\left(\psi^{(\varepsilon)}_{n,d_{n}}(Z,\overline{Z}),\ .\ \right)_{\fH_{f}}, \\ \nonumber
\sum_{n}\hat{\Pi}_{n} & = \mathds{1}_{L^{2}(\R,\fH_{f})},
\end{align}
which are expected to be almost projections in the $\cO(\varepsilon)$-sense, because the coherent state family $\{\zeta^{(\varepsilon)}_{Z}\}_{Z\in\C^{d}}$ becomes orthogonal in this limit:
\begin{align}
\label{eq:almostprojection}
(\hat{\Pi}^{\varepsilon}_{n})^{2} = \hat{\Pi}^{\varepsilon}_{n} + \cO(\varepsilon).
\end{align}
On the same grounds, the electronic energy and the almost projections commute in the $\cO(\varepsilon)$-sense,
\begin{align}
\label{eq:electronicalmostinvariance}
[H_{e}(\hat{Q}_{\varepsilon},\hat{P}_{\varepsilon};\hat{q},\hat{p}),\hat{\Pi}^{\varepsilon}_{n}]=\cO(\varepsilon),
\end{align}
and we may hope for (cp. \eqref{eq:firstorderinvariantsubspace}):
\begin{align}
\label{eq:almostinvariantsubspacefirstorder}
[\hat{H}_{\mol},\hat{\Pi}^{\varepsilon}_{n}] & = \cO(\varepsilon)
\end{align}
on a suitable domain of bounded energy states.\\[0.1cm]
Regarding the computation of effective Hamiltonians for the (almost) subspaces $\img\hat{\Pi}^{\varepsilon}_{n}$ in the sense of \eqref{eq:moleculareffectiveHamiltonian}, there is another catch, due to the coherent states not being orthogonal: In the conventional Born-Oppenheimer ansatz the expression for the effective Hamiltonian can be obtained from the restriction of \eqref{eq:moleculareigenvalueprojection} to the n-th electronic band. This is no longer the case in the coherent state framework. While the explicit expression for the effective Hamiltonian, $\hat{H}^{\eff,\varepsilon}_{\mol,n} = \Pi^{\varepsilon}_{n}\hat{H}_{\mol}\Pi^{\varepsilon}_{n}$, becomes more involved because of the absence of orthogonality relations, the analogue of \eqref{eq:moleculareigenvalueprojection} is still rather simple ($A=0$),
\begin{align}
\label{eq:coherentstateeigenvalueprojection}
 & E\ \Psi^{E,(\varepsilon)}_{n,d_{n}}(\overline{Z};Z,\overline{Z}) \\ \nonumber
 & = \sum_{\substack{ n',d_{n'} \\ n'',d_{n''} }}\!\!\left(\!\varepsilon\left(\overline{\partial}_{\sA}\right)^{n',d_{n'}}_{n,d_{n}}-\frac{1}{2}\overline{Z}\delta_{n,n'}\delta_{d_{n},d_{n'}}\!\right)\!\cdot\!\left(\!\varepsilon\left(\overline{\partial}_{\sA}\right)^{n'',d_{n''}}_{n',d_{n'}}-\frac{1}{2}\overline{Z}\delta_{n',n''}\delta_{d_{n'},d_{n''}}\!\right)\Psi^{E,(\varepsilon)}_{n'',d_{n''}}(\overline{Z};Z,\overline{Z}) \\ \nonumber
 &\hspace{0.5cm}+\int_{\C^{d}}\frac{d^{2}Z'}{(\varepsilon\pi)^{d}}\sum_{n',d_{n'}}e_{n,\varepsilon}(Z',\overline{Z'})\sK^{(\varepsilon)}(\overline{Z},Z';(Z,\overline{Z}),(Z',\overline{Z'}))^{n',d_{n'}}_{n,d_{n}}\Psi^{E,(\varepsilon)}_{n',d_{n'}}(\overline{Z'};Z',\overline{Z'}),
\end{align}
where we introduced the following objects:
\begin{align}
\Psi^{E,(\varepsilon)}_{n,d_{n}}(\overline{Z};Z,\overline{Z}) & = \left(\zeta^{(\varepsilon)}_{Z}\otimes\psi^{(\varepsilon)}_{n,d_{n}}(Z,\overline{Z}),\Psi^{E}\right)_{L^{2}(\R^{d},\fH_{f})}, \\ \nonumber
\left(\overline{\partial}_{\sA}\right)^{n',d_{n'}}_{n,d_{n}} & = \delta_{n,n'}\delta_{d_{n},d_{n'}}\nabla_{\overline{Z}}-i\overline{\sA}(Z,\overline{Z})^{n',d_{n'}}_{n,d_{n}}, \\ \nonumber
\overline{\sA}(Z,\overline{Z})^{n',d_{n'}}_{n,d_{n}} & = i\left(\psi^{(\varepsilon)}_{n,d_{n}}(Z,\overline{Z}),\nabla_{\overline{Z}}\ \psi^{(\varepsilon)}_{n',d_{n'}}(Z,\overline{Z})\right)_{\fH_{f}}, \\ \nonumber
\sK^{(\varepsilon)}(\overline{Z},Z';(Z,\overline{Z}),(Z',\overline{Z'}))^{n',d_{n'}}_{n,d_{n}} & = (\zeta^{(\varepsilon)}_{Z},\zeta^{(\varepsilon)}_{Z'})_{L^{2}(\R^{d})}(\psi^{(\varepsilon)}_{n,d_{n}}(Z,\overline{Z}),\psi^{(\varepsilon)}_{n',d_{n'}}(Z',\overline{Z'}))_{\fH_{f}}.
\end{align}
Here, $\sK^{(\varepsilon)}$ is a reproducing kernel in the \textit{Segal-Bargmann space} $\sS\!\sB^{2}_{\varepsilon}(\C^{d},\fH_{f})$ defined by the coherent states $\zeta^{(\varepsilon)}_{Z},\ Z\in\C^{d}$:
\begin{align}
\label{eq:segalbargmannkernel}
\Psi^{(\varepsilon)}_{n,d_{n}}(\overline{Z};Z,\overline{Z}) & = \int_{\C^{d}}\frac{d^{2}Z'}{(\varepsilon\pi)^{d}}\sK^{(\varepsilon)}(\overline{Z},Z';(Z,\overline{Z}),(Z',\overline{Z'}))^{n',d_{n'}}_{n,d_{n}}\sum_{n',d_{n'}}\Psi^{(\varepsilon)}_{n',d_{n'}}(\overline{Z'};Z',\overline{Z'}),
\end{align}
and $\overline{\sA}$ is (a part of) a phase space generalisation (in complex $(Z,\overline{Z})$-coordinates) of the Berry-Simon connection (cp. \eqref{eq:molecularpregaugefield}). Since the coherent states $\zeta^{(\varepsilon)}_{Z},\ Z\in\C^{d},$ are holomorphic up to a factor $e^{-\frac{1}{2\varepsilon}Z\cdot\overline{Z}}$, \eqref{eq:coherentstateeigenvalueprojection} has to be supplemented by a ``flatness constrain'', which characterises elements of $\sS\!\sB^{2}_{\varepsilon}(\C^{d},\fH_{f})$ (apart from $L^{2}$-integrability):
\begin{align}
\label{eq:flatnessconstraint}
\sum_{n',d_{n'}}\varepsilon\underbrace{\left(\delta_{n,n'}\delta_{d_{n},d_{n'}}\nabla_{Z}-i\sA(Z,\overline{Z})^{n',d_{n'}}_{n,d_{n}}\right)}_{:=\left(\partial_{\sA}\right)^{n',d_{n'}}_{n,d_{n}}}\Psi^{E,(\varepsilon)}_{n',d_{n'}}(\overline{Z};Z,\overline{Z}) = -\frac{1}{2}Z\Psi^{E,(\varepsilon)}_{n,d_{n}}(\overline{Z};Z,\overline{Z}).
\end{align}
Related to this aspect of the coherent state ansatz is the observation that the (upper) symbol of a coherent state de-quantized operator is not immediately obtained from the (diagonal) expectation values of the operator w.r.t. the coherent states, i.e. the \textit{lower symbol}. Instead, we find:
\begin{align}
\label{eq:uppertolowersymbol}
 &  L^{\varepsilon}_{f}(Z,\overline{Z})^{n,d_{n}}_{n',d_{n'}} \\ \nonumber
 & = \left(\zeta^{(\varepsilon)}_{Z}\otimes\psi^{(\varepsilon)}_{n,d_{n}}(Z,\overline{Z}),f(\hat{A}_{\varepsilon},\hat{A}^{*}_{\varepsilon};\hat{q},\hat{p})\zeta^{(\varepsilon)}_{Z}\otimes\psi^{(\varepsilon)}_{n',d_{n'}}(Z,\overline{Z})\right)_{L^{2}(\R^{d},\fH_{f})} \\ \nonumber
 & = \int_{\C^{d}}\frac{d^{2}Z'}{(\varepsilon\pi)^{d}}\sum_{\substack{ n'',d_{n''} \\ n''',d_{n'''} }}\!\!\!\!\left|(\zeta^{(\varepsilon)}_{Z},\zeta^{(\varepsilon)}_{Z'})_{L^{2}(\R^{d})}\right|^{2}f(Z,\overline{Z};\hat{q},\hat{p})^{n'',d_{n''}}_{n''',d_{n'''}} \\[-0.75cm] \nonumber
 &\hspace{3.5cm}\times (\psi^{(\varepsilon)}_{n,d_{n}}(Z,\overline{Z}),\psi^{(\varepsilon)}_{n'',d_{n''}}(Z',\overline{Z'}))_{\fH_{f}}(\psi^{(\varepsilon)}_{n''',d_{n'''}}(Z',\overline{Z'}),\psi^{(\varepsilon)}_{n',d_{n'}}(Z,\overline{Z}))_{\fH_{f}}.
\end{align}
Therefore, the task of computing the upper symbol of the effective Hamiltonian, as a candidate for a generator of the semi-classical dynamics of the slow degrees of freedom, is a rather cumbersome task, and might not even be possible. Moreover, while the lower symbol of an operator is unambigously defined by the coherent states (if it exists), and can even be argued to characterise the operator uniquely for certain kinds of \textit{complete families of coherent states} (not to be confused with completeness in the Hilbert basis sense\cite{SimonTheClassicalLimit, KlauderCoherentStatesApplications}), the upper symbol will only be unique, if a suitable class of symbols is identified. The dichotomy between upper and lower symbol can be further exemplified by their duality regarding the trace:
\begin{align}
\label{eq:traceduality}
\tr_{L^{2}(\R^{d},\fH_{f})}\left(f(\hat{A}_{\varepsilon},\hat{A}^{*}_{\varepsilon})\hat{F}\right) & = \int_{\C^{d}}\frac{d^{2}Z}{(\varepsilon\pi)^{d}}\tr_{\fH_{f}}\left(f(Z,\overline{Z})L^{\varepsilon}_{F}(Z,\overline{Z})\right),
\end{align}
for $f\in L^{1}(\C^{d},\cS_{1}(\fH_{f}))$ and $\hat{F}\in\cB(L^{2}(\R^{d},\fH_{f}))$\cite{SimonTheClassicalLimit}.\\[0.25cm]
Let us now turn to the apparent questions and problems the sketched approach raises.\\[0.1cm]
Firstly, we may ask, why we should only subject the electronic part of $\hat{H}_{\mol}$ to the coherent state (de-)quantisation, as this is not forced upon us, in comparison with the conventional Born-Oppenheimer ansatz, where the range of the pure state quantisation associated with the mutually commuting coupling operators $Q_{i},\ i=1,...,d,$ is severely restricted by orthogonality relations. Secondly, we should wonder, how we are supposed to obtain a systematic perturbation theory of the involved $\varepsilon$-dependent objects, as already the leading order approximation of the restricted dynamics (to $\img\hat{\Pi}^{\varepsilon}_{n}$) presumably requires control of the first order $\varepsilon$-expansion of the effective Hamiltonian. Thirdly, we might want to achieve a more symmetric form of (de-)quantisation, especially regarding \eqref{eq:traceduality}.\\[0.1cm]
In view of the next subsection and the fact that we should not expect a splitting, as in the example at hand, for Hamiltonians describing more general systems, the answer to the first question is that we should not do so (a very instructive example is provided by the Dirac equation with slowly varying potentials\cite{PanatiSpaceAdiabaticPerturbation}).\\
Addressing the second problem is slightly more subtle, but a practicable answer, also justified by the successes of adiabatic perturbation theory, is that we should not only look for a phase space quantisation, but a \textit{deformation quantisation} with sufficiently broad range to cover interesting operators, i.e. we would like to have a de-quantisation of a large enough (depending on the problem) algebra of operators on the total Hilbert space ($L^{2}(\R^{d},\fH_{f})$ in our example) such that the operator product (in the slow sector) gets mapped to a non-commutative ($\star_{\varepsilon}$-)product on a suitable class of phase space functions (with values in operators on $\fH_{f}$). Moreover, to be able to control the errors, arising in the perturbative expansion in $\varepsilon$, a certain notion of continuity of the quantisation should be available. The third point, can dealt with by replacing the coherent state quantisation with the Weyl quantisation, and turns out to be intimately connected with the second point.
\section{Weyl quantisation and space adiabatic perturbation theory}
\label{sec:weyl}
The issues raised at the end of the previous subsection, can be addressed in the setting of space adiabatic perturbation theory, which was developed by Panati, Spohn and Teufel in \cite{PanatiSpaceAdiabaticPerturbation}\footnote{Cf. also \cite{PanatiEffectiveDynamicsFor}, and \cite{TeufelAdiabaticPerturbationTheory} for a comprehensive review}.\\
We recall in this section the main steps and ideas of this program, and what are the necessary ingredients to implement them. The original work of Panati, Spohn and Teufel is phrased in terms of (equivariant) Weyl quantisation, when the phase space of the slow variables can be realised as $T^{*}\R^{d}\cong\R^{2d}$ (or a quotient thereof by a lattice $\Gamma\subset\R^{d}$ in the equivariant case).\\[0.1cm]
To begin with, the quantum dynamical system, $(\fH,(\hat{H},D(\hat{H})))$, to be considered, given in terms of a (self-adjoint) Hamiltonian $\hat{H}$ acting on a (dense) domain $D(\hat{H})$ in Hilbert space $\fH$, should admit the following general description:
\begin{itemize}
	\item[(a)] There is a splitting of the Hilbert space, $\fH$, into slow, $\fH_{s}$, and fast, $\fH_{f}$, degrees of freedom. The separation the two sectors is controlled by a (small) dimensionless parameter $\varepsilon$.
	\item[(b)] There is a (continuous) deformation quantisation (with deformation parameter $\varepsilon$),
\begin{align}
\label{eq:generaldeformationquant}
\widehat{\ .\ }^{\varepsilon}\ :\ & S^{\infty}(\varepsilon;\Gamma,\cB(\fH_{f}))\subset C^{\infty}(\Gamma,\cB(\fH_{f}))\longrightarrow L(\fH),
\end{align}
of the (classical) phase space, $\Gamma$, of the slow variables with values in linear operators, $L(\fH)$, on $\fH$. Here, $S^{\infty}(\varepsilon;\Gamma,\cB(\fH_{f}))$ is a class of ($\varepsilon$-dependent) quantisable functions, the \textit{semi-classical symbols}, on $\Gamma$ with values in bounded operators, $\cB(\fH_{f})$, on $\fH_{f}$. The operator product in $L(\fH)$ is reflected in a (continuous) $\star_{\varepsilon}$-product on $S^{\infty}(\varepsilon;\Gamma,\cB(\fH_{f}))$. Elements of the latter admit an asymptotic expansion in $\varepsilon$, increasing the regularity (boundedness) of (operator-)contributions with increasing order, and compatible with a similar expansion of $\star_{\varepsilon}$ (\textit{Moyal product}). Quantisations of $\cO(\varepsilon^{\infty})$-elements in $S^{\infty}(\varepsilon;\Gamma,\cB(\fH_{f}))$ are ``small'' bounded operators (\textit{smoothing operators}). Moreover, the quantisation encompasses the Hamiltonian $\hat{H}$, i.e. there is a semi-classical symbol $H_{\varepsilon}\in S^{\infty}(\varepsilon;\Gamma,\cB(\fH_{f}))$ (taking values in self-adjoint operators on $\fH_{f}$\footnote{This is a typical feature of Weyl and coherent state quantisation, in contrast to Kohn-Nirenberg quantisation\cite{FollandHarmonicAnalysisIn}. The technical advantage of self-adjoint symbols is a better control of their spectral properties, when $\fH_{f}$ is infinite-dimensional.}) with asymptotic expansion
\begin{align}
\label{eq:generalHamiltonianasymptotics}
H_{\varepsilon} & \sim \sum^{\infty}_{k=0}\varepsilon^{k}H_{k},\ \forall k\in\N_{0}: H_{k}\in C^{\infty}(\Gamma,\cB(\fH_{f})),
\end{align}
such that $\widehat{H_{\varepsilon}}^{\varepsilon} = \hat{H}$.
	\item[(c)] There is a relevant part, $\sigma_{*}(H_{0})$, of the (point-wise) spectrum, $\sigma(H_{0})=\{\sigma(H_{0}(\gamma))\}_{\gamma\in\Gamma}$, of the \textit{principal symbol} $H_{0}$, that is isolated from the (point-wise) remainder, $\sigma^{c}_{*}(H_{0})$, by a finite gap (global over $\Gamma$)\footnote{Precise conditions characterising the gap need to be adapted to the type of Hamiltonian $\hat{H}$\cite{TeufelAdiabaticPerturbationTheory}.}.
\end{itemize}
Let us briefly explain the meaning the three conditions just stated:\\[0.1cm]
Clearly, (a) describes the identification of the (two) sectors of the quantum system, which are to be considered as being separated by different time scales w.r.t. the dynamics ($\varepsilon$ quantifies the separation).\\
(b) provides a kind of minimal (formal) framework, that is necessary to establish a systematic perturbation theory that exploits the separation of scales defined by (a) (adiabatic perturbation theory in orders of $\varepsilon$). Thus, the existence of an appropriate deformation (de-)quantisation procedure (and a compatible symbolic calculus) to handle operators in the slow sectors is the main technical building block, upon which the whole program of space adiabatic perturbation theory rests.\\
(c) defines the starting point of the perturbation theory in the limit of infinitely separated times scales ($\varepsilon\rightarrow0$, frozen dynamics of the slow system). That is, the spectral problem of the fast variables is assumed to be under sufficient control for fixed (classical) states of the slow variables, and is used as input for an analysis of the dynamics of the coupled system (this is analogous to electronic structure calculations in the conventional Born-Oppenheimer approach). A gap, isolating an interesting part, $\sigma_{*}(H_{0})$, of the spectrum of $H_{0}$, is typically necessary to control the error in perturbation theory coming from non-adiabatic transitions, as these are dealt with by bounds on the (local) resolvents of $\widehat{H_{0}}^{\varepsilon}$ w.r.t. $\sigma_{*}(H_{0})$). \\[0.25cm] 
Assuming that the above conditions are satisfied, the program consists roughly of four steps:
\begin{itemize}
	\item[1.] Denoting the (smooth\footnote{This should be implied by the gap condition.}) spectral projection of $H_{0}$ onto the relevant part $\sigma_{*}$ by $\pi_{0}$, an almost invariant projection $\hat{\Pi}^{\varepsilon}$, i.e.
\begin{align}
\label{eq:almostinvariantprojection}
[\hat{H},\hat{\Pi}^{\varepsilon}] & = \cO_{0}(\varepsilon^{\infty}),
\end{align}
is constructed, such that $\hat{\Pi}^{\varepsilon} = \widehat{\pi_{\varepsilon}}^{\varepsilon}+\cO_ {0}(\varepsilon^{\infty})$ is close to the quantisation of a (bounded) semi-classical symbol $\pi_{\varepsilon}\in S^{\infty}(\varepsilon;\Gamma,\cB(\fH_{f}))$. The notation $\cO_{0}(\varepsilon^{\infty})$ indicates that the operator norm of the left hand side should be bounded by any power of $\varepsilon$ (uniformly for $\varepsilon\in(0,\varepsilon_{0}]$ for some $\varepsilon_{0}>0$\cite{TeufelAdiabaticPerturbationTheory}.\\
The semi-classical symbol $\pi_{\varepsilon}$ has an asymptotic expansion with principal symbol $\pi_{0}$,
\begin{align}
\label{eq:almostinvariantprojectionasymptotics}
\pi_{\varepsilon} & \sim \sum^{\infty}_{k=0}\varepsilon^{k}\pi_{k},
\end{align}
that qualifies as an invariant projection relative to Moyal product (characterising it uniquely), i.e.
\begin{align}
\label{eq:moyalprojection}
\pi_{\varepsilon}\star_{\varepsilon}\pi_{\varepsilon} & = \pi_{\varepsilon}, & \pi^{*}_{\varepsilon} & = \pi_{\varepsilon}, & [H_{\varepsilon},\pi_{\varepsilon}]_{\star_{\varepsilon}} & = 0.
\end{align}
The subspace $\hat{\Pi}^{\varepsilon}\fH\subset\fH$ is called \textit{almost invariant subspace}, as it remains approximately invariant w.r.t. the dynamics (\textit{Duhamel's formula}):
\begin{align}
\label{eq:almostinvariantprojectiondynamics}
[e^{-i\hat{H}s},\hat{\Pi}^{\varepsilon}] & = \cO_{0}(|s|\varepsilon^{\infty}).
\end{align}
The limit $\varepsilon\rightarrow0$ is not necessarily meaningful for $\hat{\Pi}^{\varepsilon}\fH$.
	\item[2.] Next, a unitary operator $\hat{U}^{\varepsilon}\in\cB(\fH)$ is constructed, that identifies the almost invariant subspace $\hat{\Pi}^{\varepsilon}\fH$ with an $\varepsilon$-independent reference (sub)space $\hat{\Pi}_{r}\fH$, which allows for a simple description. Similar to $\hat{\Pi}^{\varepsilon}$, $\hat{U}^{\varepsilon}$ is $\cO(\varepsilon^{\infty})$-close to the quantisation of a semi-classical symbol $u_{\varepsilon}\in S^{\infty}(\varepsilon;\Gamma,\cB(\fH_{f}))$. The latter has an asymptotic expansion,
\begin{align}
\label{eq:moyalunitaryasymptotics}
u_{\varepsilon} & \sim \sum^{\infty}_{k=0}\varepsilon^{k}u_{k},
\end{align}
with a (smooth) unitary-valued principal symbol $u_{0}$, that defines a \textit{reference projection} $\pi_{r}\in\cB(\fH_{f})$ by
\begin{align}
\label{eq:referenceequation}
u_{0}(\gamma)\pi_{0}(\gamma)u_{0}(\gamma)^{*} & = \pi_{r}.
\end{align}
$u_{0}$ is called the \textit{reference unitary}. The quantisation, $\hat{\Pi}_{r}=\mathds{1}_{\fH_{s}}\otimes\pi_{r}$, of the reference projection $\pi_{r}$ defines the reference space $\hat{\Pi}_{r}\fH$. It holds, that:
\begin{align}
\label{eq:almostinvariantunitary}
\hat{\Pi}_{r} & = \hat{U}^{\varepsilon}\hat{\Pi}^{\varepsilon}(\hat{U}^{\varepsilon})^{*}.
\end{align}
The asymptotic expansion of $u_{\varepsilon}$ is characterised (although not uniquely) by the following properties w.r.t. the Moyal product:
\begin{align}
\label{eq:moyalunitary}
u_{\varepsilon}\star_{\varepsilon}u^{*}_{\varepsilon} & = 1 = u^{*}_{\varepsilon}\star_{\varepsilon}u_{\varepsilon}, & u_{\varepsilon}\star_{\varepsilon}\pi_{\varepsilon}\star_{\varepsilon}u^{*}_{\varepsilon} & = \pi_{r}.
\end{align}
Clearly, $u_{0}$ gives a (global) trivialisation of the \textit{adiabatic bundle} $\pi_{0}\fH\rightarrow\Gamma$.
	\item[3.] In a third step, the dynamics generated by $\hat{H}$ (almost) inside $\hat{\Pi}^{\varepsilon}\fH$ is mapped to the reference space $\hat{\Pi}_{r}\fH$, where it is generated by a (self-adjoint) effective Hamiltonian $\hat{h}$. Due to the fact that $\hat{\Pi}^{\varepsilon}$ and $\hat{U}^{\varepsilon}$ are $\cO_{0}(\varepsilon^{\infty})$-close to quantisations of a Moyal projection $\pi_{\varepsilon}$ and a Moyal unitary $u_{\varepsilon}$, respectively, it is possible to define $\hat{h}$ as the quantisation of a self-adjoint semi-classical symbol $h_{\varepsilon}\in S^{\infty}(\varepsilon;\Gamma,\cB(\fH_{f}))$, the \textit{effective Hamiltonian symbol}:
\begin{align}
\label{eq:effectiveHamniltoniandefinition}
h_{\varepsilon} & \sim u_{\varepsilon}\star_{\varepsilon}H_{\varepsilon}\star_{\varepsilon}u^{*}_{\varepsilon}.
\end{align}
or
\begin{align}
\label{eq:effectiveHamniltoniandefinitionalternative}
\pi_{r}\star_{\varepsilon}h_{\varepsilon}\star_{\varepsilon}\pi_{r} & \sim \pi_{r}\star_{\varepsilon}u_{\varepsilon}\star_{\varepsilon}H_{\varepsilon}\star_{\varepsilon}u^{*}_{\varepsilon}\star_{\varepsilon}\pi_{r} \\ \nonumber
 & \sim u_{\varepsilon}\star_{\varepsilon}\pi_{\varepsilon}\star_{\varepsilon}H_{\varepsilon}\star_{\varepsilon}\pi_{\varepsilon}\star_{\varepsilon}u^{*}_{\varepsilon},
\end{align}
which makes a computation of the $\cO(\varepsilon^{n})$-truncations $h_{\varepsilon,(n)}$ and $\pi_{r}\star_{\varepsilon}h_{\varepsilon,(n)}\star_{\varepsilon}\pi_{r}$ computationally feasible. The effective Hamiltonian $\hat{h}$ satisfies by construction:
\begin{align}
\label{eq:spaceadiabaticcondition}
[\hat{h},\hat{\Pi}_{r}] & = 0 \\ \nonumber
e^{-i\hat{H}s}-(\widehat{u_{\varepsilon}}^{\varepsilon})^{*}e^{-i\hat{h}s}\widehat{u_{\varepsilon}}^{\varepsilon} & = \cO_ {0}(|s|\varepsilon^{\infty}), \\ \nonumber
e^{-i\hat{H}s}-(\hat{U}^{\varepsilon})^{*}e^{-i\hat{h}s}\hat{U}^{\varepsilon} & = \cO_{0}((1+|s|)\varepsilon^{\infty}),
\end{align}
which entails the \textit{space adiabatic theorem with time scale} $t>0$:
\begin{align}
\label{eq:spaceadiabatictheorem}
e^{-i\hat{H}s}\hat{\Pi}^{\varepsilon}-(\widehat{u_{\varepsilon,(n)}}^{\varepsilon})^{*}e^{-i\widehat{h_{\varepsilon,(n+k)}}^{\varepsilon}s}\hat{\Pi}_{r}\widehat{u_{\varepsilon,(n)}}^{\varepsilon} & = \cO_{0}((1+|t|)\varepsilon^{\infty}),
\end{align}
for large enough $n,k\in\N_{0}, |s|\leq\varepsilon^{-k}t$. Here, $\widehat{u_{\varepsilon,(n)}}^{\varepsilon}$ and $\widehat{h_{\varepsilon,(n+k)}}^{\varepsilon}$ are the quantisations of the $\cO(\varepsilon^{n+1})$- and $\cO(\varepsilon^{n+k+1})$-truncation of \eqref{eq:moyalunitaryasymptotics} and \eqref{eq:effectiveHamniltoniandefinition}, respectively.\\
The meaning of \eqref{eq:spaceadiabatictheorem} is that the error of the adiabatic approximation is controlled, not only, by the order of the expansion of the effective Hamiltonian, but, to the same extent, by the order of the expansion of the reference projection $\hat{\Pi}^{\varepsilon}$ and its associated unitary $\hat{U}^{\varepsilon}$. Solely expanding the effective Hamiltonian at a given level of the error, improves the time scale on which the adiabatic approximation remains valid.\\
The requirement to choose $n,k\in\N_{0}$ large enough in the expansions of $\widehat{u_{\varepsilon}}^{\varepsilon}$ and $\widehat{h_{\varepsilon}}^{\varepsilon}$ is necessary, because the leading orders of $\widehat{h_{\varepsilon}}^{\varepsilon}$ are typically unbounded upon quantisation. 
	\item[4.] Finally, if $\sigma_{*}(H_{0})=\{E_{*}\}$ consists of a single (possibly degenerate) eigenvalue, a \textit{semi-classical approximation} to the dynamics inside the almost invariant subspace can be made. From \textit{Heisenberg's equation} on the reference space $\hat{\Pi}_{r}\fH$,
\begin{align}
\label{eq:referenceheisenbergequation}
\widehat{O_{\varepsilon}}^{\varepsilon}(t) & = e^{\frac{i}{\varepsilon}\hat{h}t}\widehat{O_{\varepsilon}}^{\varepsilon}e^{-\frac{i}{\varepsilon}\hat{h}t},\ \ \ \left(\partial_{t}\widehat{O_{\varepsilon}}^{\varepsilon}\right)(t)  = \frac{i}{\varepsilon}[\hat{h},\widehat{O_{\varepsilon}}^{\varepsilon}(t)],
\end{align}
\textit{Egorov's hierachy} is derived for semi-classical observables $O_{\varepsilon}\in S^{\infty}(\varepsilon;\Gamma,\cB(\pi_{r}\fH_{f}))$ (expansion on the level of symbols):
\begin{align}
\label{eq:egorovhierachy0}
\Rightarrow\hspace{1cm} \left(\partial_{t}O_{0}\right)(t) & = \{E_{*},O_{0}(t)\} + i[h_{1},O_{0}(t)] \\
\label{eq:egorovhierachy1}
 \left(\partial_{t}O_{1}\right)(t) & = \{E_{*},O_{1}(t)\} + i[h_{1},O_{1}(t)] + \tfrac{1}{2}\left(\{h_{1},O_{0}(t)\}-\{O_{0}(t),h_{1}\}\right) \\ \nonumber
 & \hspace{0.5cm}+ i[h_{2},O_{0}(t)] \\ \nonumber
 & \vdots
\end{align}
The latter can be solved for $O_{n},\ n\in\N_{0}$ iteratively, because the equation at the $n$-th level only depends on $O_{m},\ m\leq n$. The solution at the lowest order, \eqref{eq:egorovhierachy0}, i.e. the time evolution of the principal symbol $O_{0}$, is determined by the classical flow generated by the Hamiltonian function $h_{0} = E_{*}$, and, in the case of a degenerate eigenvalue, the unitary evolution generated by $h_{1}$ transported along the flow of $h_{0}$:
\begin{align}
\label{eq:egorovevolution0}
O_{0}(\gamma,t) & = V(\gamma,t)^{*}O_{0}(\Phi_{t}(\gamma))V(\gamma,t), & O_{0}(\gamma,0) & = O_{0}(\gamma),\ \gamma\in\Gamma,
\end{align}
where
\begin{align}
\label{eq:egorovevolution0flows}
\partial_{t}\Phi_{t}(\gamma) & = X_{E^{*}}(\gamma), & \partial_{t}V(\gamma,t) & = -i h_{1}(\Phi_{t}(\gamma))V(\gamma,t) \\ \nonumber
\Phi_{0}(\gamma) & = \gamma, & V(\gamma,0) & = \mathds{1}_{\pi_{r}\fH_{f}},
\end{align}
with $X_{E_{*}}$ denoting the Hamiltonian vector field of $E_{*}$ w.r.t. the symplectic structure on $\Gamma$. For scalar principal symbols $O_{0}=o_{0}\mathds{1}_{\pi_{r}\fH_{f}}$ \eqref{eq:egorovevolution0} reduces to
\begin{align}
\label{eq:egorovevolution0scalar}
o_{0}(\gamma,t) & = o_{0}(\Phi_{t}(\gamma)), & o_{0}(\gamma,0) & = o_{0}(\gamma),\ \gamma\in\Gamma,
\end{align}
hence the name semi-classical approximation.\\
Under suitable assumptions (gap conditions) the quantisation of the $\cO(\varepsilon^{n+1})$-expansion of a semi-classical observable $O_{\varepsilon}$, subject to semi-classical time-evolution, can be related to the quantum dynamics in the reference space. For example, regarding the lowest order semi-classical flow \eqref{eq:egorovevolution0flows}, a first order \textit{Egorov's theorem} for the principal part $O_{0}$ and its quantisation is conceivable\cite{PanatiSpaceAdiabaticPerturbation}:
\begin{align}
\label{eq:}
\forall T\in\R_{\geq0}:\exists C_{T}>0:\forall t\in[-T,T]:\ \left|\left|e^{\frac{i}{\varepsilon}\hat{h}t}\widehat{O_{0}}^{\varepsilon}e^{-\frac{i}{\varepsilon}\hat{h}t}-\widehat{O_{0}(t)}^{\varepsilon}\right|\right|_{\cB(\hat{\Pi}_{r}\fH)}\leq\varepsilon\ C_{T}.
\end{align}
In view of the time evolution described by \eqref{eq:referenceheisenbergequation}, it has to be kept in mind, that this equation encodes the $\cO(\varepsilon^{\infty})$-approximation of the original quantum dynamics generated by $\hat{H}$, after mapping it to the reference space $\hat{\Pi}_{r}\fH$. The upshot of this is, that semi-classical observables inside the reference space, $O_{\varepsilon}\in S^{\infty}(\varepsilon;\Gamma,\cB(\pi_{r}\fH_{f}))$, correspond (up to $\cO_{0}(\varepsilon^{\infty})$) to semi-classical observables inside the almost invariant subspace $\hat{\Pi}^{\varepsilon}\fH$\cite{TeufelAdiabaticPerturbationTheory}. More precisely, semi-classical observables w.r.t. $\hat{\Pi}^{\varepsilon}\fH$ are modelled by operators $\hat{O}\in L(\fH)$ that are almost diagonal w.r.t. $\hat{\Pi}^{\varepsilon}$:
\begin{align}
\label{eq:almostdiagonalobservables}
[\hat{O},\hat{\Pi}^{\varepsilon}] & = \cO_{0}(\varepsilon^{\infty}).
\end{align}
The dynamics of general observables $\hat{O}\in L(\fH)$ can be considered in the weak sense of restricting to expectation values w.r.t. states of the physical system in $\hat{\Pi}^{\varepsilon}\fH$. This amounts to projecting $\hat{O}$ to the almost invariant subspace:
\begin{align}
\label{eq:weak}
\hat{O}_{|\hat{\Pi}^{\varepsilon}\fH} = \hat{\Pi}^{\varepsilon}\hat{O}\hat{\Pi}^{\varepsilon}.
\end{align}
\end{itemize}
Remarkably, 3. and 4. show that the adiabatic and semi-classical limit are completely decoupled in space adiabatic perturbation theory: While the third step fully takes place in the quantum domain, and the $\varepsilon$-quantisation is merely a technical tool to control the perturbation theory (adiabatic limit), the fourth step invokes the in build semi-classical properties of the (de-)quantisation procedure to establish a connection between the classical and quantum domains (semi-classical limit).
\section{A model with non-commutative slow variables: Spin-orbit coupling}
\label{sec:spin}
\begin{figure}[h]
\centering
\includegraphics[width=\textwidth]{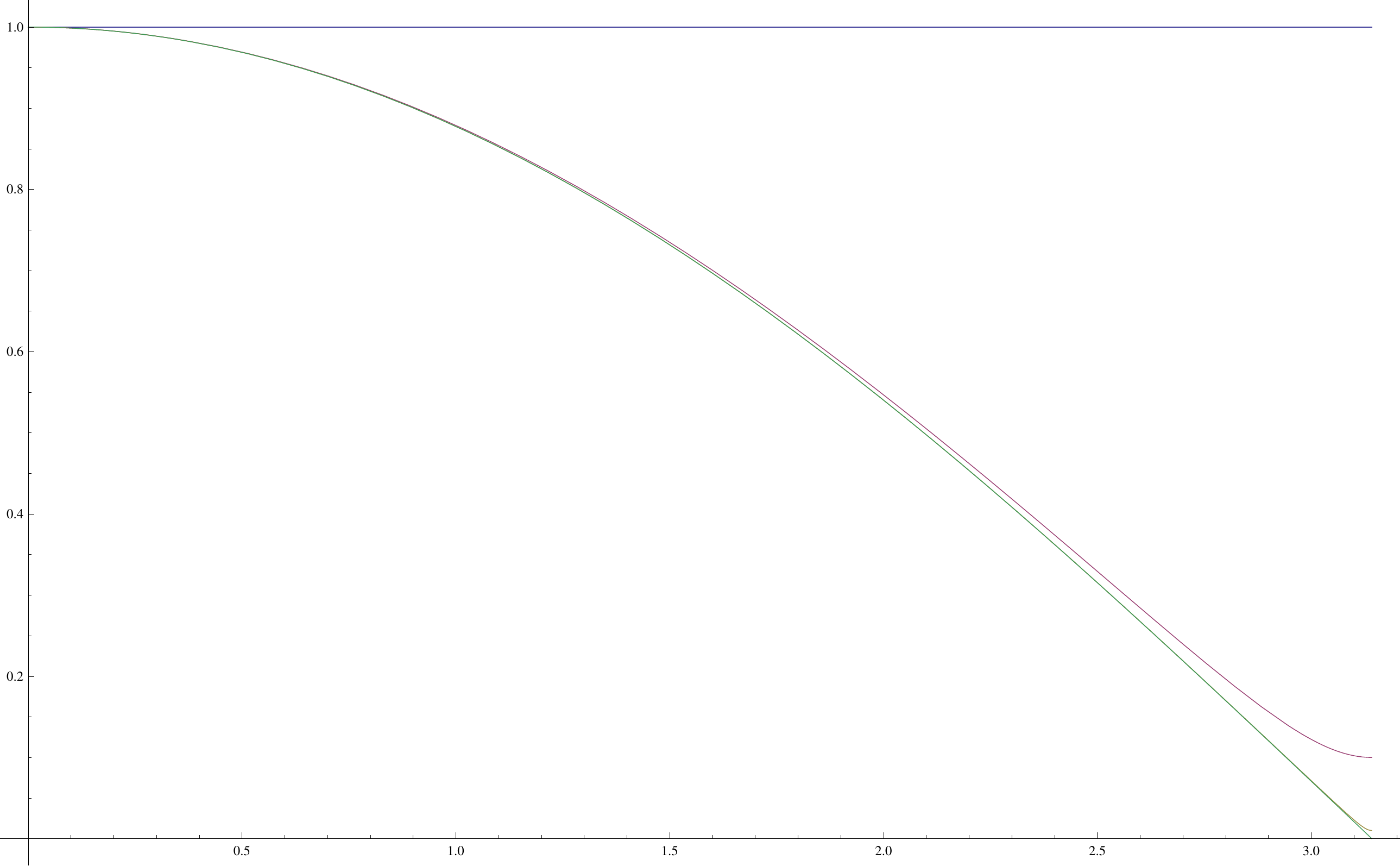}
\caption{\label{fig:spinorbitspectraldistancePauli}Plot showing the behaviour of the spectral distance $N(n,\lambda)$ as a function of $\theta\in[0,\pi)$ for $\lambda=\frac{1}{2}(1\pm10^{0}),\frac{1}{2}(1\pm10^{-1}),\frac{1}{2}(1\pm10^{-2})$ and $\frac{1}{2}$(``$=$''$\frac{1}{2}(1\pm10^{-\infty}))$ (top to bottom).}
\end{figure}

We apply the general scheme of space adiabatic perturbation theory to a simple finite dimensional model, which describes the interaction of two spin systems (\textit{spin-orbit coupling}), one of which is assumed to model the fast degrees of freedom, while the other represents the slow sector. The choice of coupled spin systems is, on the one hand, motivated by the fact that one part of the algebra of loop quantum gravity takes values in $\mathfrak{su}_{2}$\cite{ThiemannModernCanonicalQuantum, StottmeisterCoherentStatesQuantumIII}. On the other hand, the orbital angular momentum operator constitutes an easily tractable model of a vector of coupling operators, such that its components are not mutually commuting. Additionally, the model allows us to study effects of non-trivial topological structures of the \textit{adiabatic line bundles}. \\[0.25cm]
Concretely, we choose a (slightly adapted) model used by Faure and Zhilinskii \cite{FaureTopologicalPropertiesOf} to discuss the manifestation of topological indices (e.g. the Chern number) of the adiabatic line bundle, to which the Berry-Simon connection is associated, in the spectrum of the coupled system:\\[0.1cm]
We consider two spin systems ($\mathfrak{su}_{2}$-algebras), $\{J, [J_{i},J_{j}]=i\epsilon_{ijk}J_{k}\}$ and $\{S, [S_{i},S_{j}]=i\epsilon_{ijk}S_{k}\}$, (irreducibly) represented on finite-dimensional Hilbert spaces $\fH_{j}$ and $\fH_{s}$ ($d_{j}:=2j+1,d_{s}:=2s+1\in\N$, $\dim\fH_{j}=d_{j}$ and $\dim\fH_{s}=d_{s}$, $d_{j}>d_{s}$). The Hamiltonian governing the dynamics of the coupled system $\fH=\fH_{j}\otimes\fH_{s}$ is:
\begin{align}
\label{eq:spinorbitHamiltonian}
\hat{H}^{\lambda}_{d_{j}} & = (1-\lambda)\mathds{1}_{\fH_{j}}\otimes S_{3} + \lambda\frac{2}{d_{j}} J\cdot S,\ \ \ \lambda\in[0,1].
\end{align}
Faure and Zhilinskii use the pre-factor $\frac{1}{j}$, instead of $\frac{2}{d_{j}}$, in front of the coupling term $J\cdot S$. But, the factor $\frac{1}{d_{j}}$ turns out to be the expansion parameter of the $\star$-product to be introduced in this context\cite{StratonovichOnDistributionsIn, VarillyTheMoyalRepresentation, FeifeiTheWeylWigner}, and is therefore better suited for our purposes\footnote{The $\frac{1}{2}$ in the relation, $\frac{2}{d_{j}}=\frac{1}{j+\frac{1}{2}}$, between the expansion parameters can be attributed to half the sum of the positive roots of $SU(2)$\cite{SchraderSemiclassicalAsymptoticsGauge}.}. Anyway, in the (adiabatic) limit $j\rightarrow\infty$, the difference of the two factors becomes negligible.\\
For the purpose of de-quantisation of the slow sector, we employ the formalism of \textit{Stratonovich-Weyl quantisation}\footnote{Cf. \cite{VarillyTheMoyalRepresentation} for a lucid exposition of the detail with applications to spin system}, which can be summarised as follows\footnote{We reconsider this formalism in the context of semi-simple compact Lie groups \cite{FigueroaMoyalQuantizationWith} in our companion article \cite{StottmeisterCoherentStatesQuantumII}.}:
\begin{itemize}
	\item[1.] There is a $\cB(\fH_{j})$-valued function $\Delta^{j}$ on the Poisson manifold $S^{2}=\{n\in\R^{3}\ |\ n^{2}=1\}$ (two sphere), which can be used to quantise functions $f\in C^{\infty}(S^{2})$ via the formula:
	\begin{align}
	\label{eq:SWquantspin}
	\hat{A}^{d_{j}}_{f} & = \frac{d_{j}}{4\pi}\int_{S^{2}}d^{2}n\ f(n)\Delta^{j}(n)\ \in\cB(\fH_{j}),
	\end{align}
	where $d^{2}n$ denotes the $4\pi$-normalised surface measure on $S^{2}$.
	\item[2.] The de-quantisation $A^{\SW}_{d_{j}}\in C^{\infty}(S^{2})$ of an operator $\hat{A}\in\cB(\fH_{j})$, called the \textit{Stratonovich-Weyl symbol} (for short: symbol), is achieved by
	\begin{align}
	\label{eq:SWdequantspin}
	A^{\SW}_{d_{j}}(n) & = \tr_{\fH_{j}}\left(\Delta^{j}(n)\hat{A}\right).
	\end{align}
	\item[3.] $\Delta^{j}$ has the properties:
	\begin{itemize}
		\item[(a)] $\forall n\in S^{2}:\Delta^{j}(n)^{*}=\Delta^{j}(n)$,
		\item[(b)] $\frac{d_{j}}{4\pi}\int_{S^{2}}d^{2}n\ \Delta^{j}(n) = \mathds{1}_{\fH_{j}}$,
		\item[(c)] $\forall n\in S^{2}:\ \frac{d_{j}}{4\pi}\int_{S^{2}}d^{2}m\ \tr_{\fH_{j}}\left(\Delta^{j}(m)\Delta^{j}(n)\right)\Delta^{j}(m) = \Delta^{j}(n)$,
		\item[(d)] $\forall A,B\in\cB(\fH_{j}):\ \tr_{\fH_{j}}\left(\hat{A}\hat{B}\right) = \frac{d_{j}}{4\pi}\int_{S^{2}}d^{2}n\ A^{\SW}_{d_{j}}(n)B^{\SW}_{d_{j}}(n)$,
		\item[(e)] $\forall g\in SU(2): \pi_{j}(g)\Delta^{j}(n)\pi_{j}(g)^{*} = \Delta^{j}(Ad_{g}(n))$,
	\end{itemize}
	where $\pi_{j}:SU(2)\rightarrow\cB(\fH_{j})$ is the irreducible representation of $SU(2)$ defining $\fH_{j}$, and $Ad:SU(2)\rightarrow SO(3)$ is the adjoint action of $SU(2)$ under the identification $\mathfrak{su}_{2}\cong(\R^{3},\times)$.
	\item[4.] The $\star$-product of two symbols $A^{\SW}_{d_{j}}$ and $B^{\SW}_{d_{j}}$ is given by:
	\begin{align}
	\label{eq:moyalproductspin}
	\left(A^{\SW}_{d_{j}}\star B^{\SW}_{d_{j}}\right)(n) & = \frac{d_{j}}{4\pi}\int_{S^{2}}d^{2}m\frac{d_{j}}{4\pi}\int_{S^{2}}d^{2}k\ \tr_{\fH_{j}}\left(\Delta^{j}(n)\Delta^{j}(m)\Delta^{j}(k)\right)A^{\SW}_{d_{j}}(m)B^{\SW}_{d_{j}}(k).
	\end{align}
\end{itemize}
It should be noted that the quantisation, $f\mapsto\hat{A}^{d_{j}}_{f}$, is not injective, because the range of the de-quantisation, $\hat{A}\mapsto A^{\SW}_{d_{j}}$, is the $d^{2}_{j}$-dimensional subalgebra $C^{\infty}_{d_{j}}(S^{2})$ of $C^{\infty}(S^{2})$ generated by spherical harmonics $\{Y_{lm}\}_{l\in\N_{0},m=-l,..,l}\subset C^{\infty}(S^{2})$ with $l\leq 2j$. This observation is in accordance with the fact that the spherical-harmonic tensor operators $\{\hat{\mathcal{Y}}_{lm}\}_{l\in\N_{0},m=-l,..,l}$ constitute a basis for $\cB(\fH_{j})$\cite{FeifeiTheWeylWigner}.\\
Nevertheless, the $d_{j}$-expansion of the $\star$-product is computed w.r.t. $C^{\infty}(S^{2})=\bigcup_{d_{j}\in\N}C^{\infty}_{d_{j}}(S^{2})$, since the quantisation \eqref{eq:SWquantspin} projects out contribution from spherical harmonics with $l>2j$ anyway. The $d_{j}$-expansion of the $\star$-product \eqref{eq:moyalproductspin} to order $\cO(d^{-2}_{j})$ follows from the techniques in \cite{FeifeiTheWeylWigner}, although the first order expansion displayed there is incorrect (see equation (53)):
\begin{align}
\label{eq:moyalproductexpansionspin}
 & \left(A^{\SW}_{d_{j}}\star B^{\SW}_{d_{j}}\right)(n) \\ \nonumber & \sim \left(A^{\SW}_{d_{j}}\star B^{\SW}_{d_{j}}\right)_{0}(n) + d^{-1}_{j}\left(A^{\SW}_{d_{j}}\star B^{\SW}_{d_{j}}\right)_{1}(n) + d^{-2}_{j}\left(A^{\SW}_{d_{j}}\star B^{\SW}_{d_{j}}\right)_{2}(n) + \cO(d^{-3}_{j}) \\ \nonumber
 & = A^{\SW}_{0}(n)B^{\SW}_{0}(n) \\ \nonumber
 &\hspace{0.5cm} + d^{-1}_{j}\left(A^{\SW}_{0}(n)B^{\SW}_{1}(n) + A^{\SW}_{1}(n)B^{\SW}_{0}(n) - \tfrac{1}{2}A^{\SW}_{0}(n)B^{\SW}_{0}(n)\right. \\ \nonumber
 &\hspace{0.75cm}\left. + \left((n\times\nabla_{n})^{2}A^{\SW}_{0}\right)(n)B^{\SW}_{0}(n) +A^{\SW}_{0}(n)\left((n\times\nabla_{n})^{2}B^{\SW}_{0}\right)(n) \right. \\ \nonumber
 &\hspace{0.75cm}\left. + i n\cdot\left(\left(\nabla_{n}A^{\SW}_{0}\right)\times\left(\nabla_{n}B^{\SW}_{0}\right)\right)(n)\right) \\ \nonumber
 &\hspace{0.5cm} + d^{-2}_{j}\left(A^{\SW}_{0}(n)B^{\SW}_{2}(n) + A^{\SW}_{1}(n)B^{\SW}_{1}(n) + A^{\SW}_{2}(n)B^{\SW}_{0}(n)\right. \\ \nonumber
 &\hspace{0.75cm}\left. - \tfrac{1}{2}\left((n\times\nabla_{n})^{2}A^{\SW}_{0}\right)\!(n)\!\left((n\times\nabla_{n})^{2}B^{\SW}_{0}\right)\!(n)\! +\!\tfrac{1}{4}(n\times\nabla_{n})^{2}\left(\left(\nabla_{n}A^{\SW}_{0}\right)\!\cdot\!\left(\nabla_{n}B^{\SW}_{0}\right)\right)\!(n) \right. \\ \nonumber
 &\hspace{0.75cm}\left. -\tfrac{9}{4}\left(\left(\left(\nabla_{n}(n\times\nabla_{n})^{2}A^{\SW}_{0}\right)\cdot\left(\nabla_{n}B^{\SW}_{0}\right)\right)(n) + \left(\left(\nabla_{n}A^{\SW}_{0}\right)\cdot\left(\nabla_{n}(n\times\nabla_{n})^{2}B^{\SW}_{0}\right)\right)(n)\right)\right. \\ \nonumber
 &\hspace{0.75cm}\left. -\tfrac{7}{2}\left(\left(\nabla_{n}A^{\SW}_{0}\right)\cdot\left(\nabla_{n}B^{\SW}_{0}\right)\right)(n)\right.\\ \nonumber
 &\hspace{0.75cm}\left. + \left((n\times\nabla_{n})^{2}A^{\SW}_{0}\right)(n)B^{\SW}_{1}(n) + \left((n\times\nabla_{n})^{2}A^{\SW}_{1}\right)(n)B^{\SW}_{0}(n) \right.\\ \nonumber
 &\hspace{0.75cm}\left. + A^{\SW}_{0}(n)\left((n\times\nabla_{n})^{2}B^{\SW}_{1}\right)(n) + A^{\SW}_{1}(n)\left((n\times\nabla_{n})^{2}B^{\SW}_{0}\right)(n)\right.\\ \nonumber
 &\hspace{0.75cm}\left. +i n\cdot\left(\left(\left(\nabla_{n}A^{\SW}_{0}\right)\times\left(\nabla_{n}B^{\SW}_{1}\right)\right)(n) + \left(\left(\nabla_{n}A^{\SW}_{1}\right)\times\left(\nabla_{n}B^{\SW}_{0}\right)\right)(n) \right. \right.\\ \nonumber
 &\hspace{0.75cm}\left.\left. - 6\left(\left(\nabla_{n}A^{\SW}_{0}\right)\times\left(\nabla_{n}B^{\SW}_{0}\right)\right)(n) \right.\right. \\ \nonumber
 &\hspace{0.75cm}\left.\left. + \left(\left(\nabla_{n}(n\times\nabla_{n})^{2}A^{\SW}_{0}\right)\times\left(\nabla_{n}B^{\SW}_{0}\right)\right)(n) + \left(\left(\nabla_{n}A^{\SW}_{0}\right)\times\left(\nabla_{n}(n\times\nabla_{n})^{2}B^{\SW}_{0}\right)\right)(n) \right)\right) \\ \nonumber
 &\hspace{0.5cm} + \cO(d^{-3}_{j}),
\end{align}
where $A^{\SW}_{d_{j}}\sim\sum_{k=0}^{\infty}d^{-k}_{j}A^{\SW}_{k}$ and $B^{\SW}_{d_{j}}\sim\sum_{k=0}^{\infty}d^{-k}_{j}B^{\SW}_{k}$ are \textit{semi-classical symbols}. Clearly, \eqref{eq:moyalproductexpansionspin} gives the expected behaviour in the leading order of the $\star$-commutator, i.e. the imaginary unit times the Poisson bracket on $S^{2}$:
\begin{align}
\label{eq:moyalcommutatorspin}
\left[A^{\SW}_{d_{j}},B^{\SW}_{d_{j}}\right]_{\star}(j_{cl}) & \sim i j_{cl}\cdot\left(\left(\nabla_{j_{cl}}A^{\SW}_{0}\right)\times\left(\nabla_{j_{cl}}B^{\SW}_{0}\right)\right)(j_{cl}) + \cO(d^{-2}_{j}),
\end{align}
where $j_{cl}=\frac{d_{j}}{2}n$ is the ``classical'' spin vector. In contrast to the Moyal product for $\R^{2d}$, the $\cO(d^{-2}_{j})$-contribution to the $\star$-commutator does not vanish, which can be traced back to the non-trivial geometry of $S^{2}$.\\[0.25cm]
All of the above immediately generalises to the case of $\cB(\fH_{s})$-valued symbols, due to finite dimensionality. But, we have to be cautious about the ordering of symbols, as they are operator valued, e.g. \eqref{eq:moyalcommutatorspin} only holds for scalar symbols. But, \eqref{eq:moyalproductexpansionspin} was derived without assuming commutativity of the point-wise product of symbols.
\begin{Remark}
\label{rem:berezinproductspin}
In principle, we can also define a Berezin-$\star$-product for a \textit{spin coherent state quantisation} of the scale of Poisson algebras $C^{\infty}(S^{2})=\bigcup_{d_{j}\in\N}C^{\infty}_{d_{j}}(S^{2})$, because a closed de-quantisation formula, similar to \eqref{eq:SWdequantspin}, exists\cite{KlauderCoherentStatesApplications, VarillyTheMoyalRepresentation}. A $d_{j}$-expansion of this $\star$-product is arrived at via an easy, but extremely tedious, calculation along the lines of \cite{FeifeiTheWeylWigner}. We state only the result up to $\cO(d^{-2}_{j})$, as we will not make further use of it:
\begin{align}
\label{eq:berezinproductspin}
 & \left(A^{\SW}_{d_{j}}\star B^{\SW}_{d_{j}}\right)(n) \\ \nonumber
 & \sim A^{\SW}_{0}(n)B^{\SW}_{0}(n) \\ \nonumber
 &\hspace{0.5cm} + d^{-1}_{j}\left(A^{\SW}_{0}(n)B^{\SW}_{1}(n)\! +\! A^{\SW}_{1}(n)B^{\SW}_{0}(n)\! -\! \tfrac{1}{2}A^{\SW}_{0}(n)B^{\SW}_{0}(n)\! -\! \left(\left(\nabla_{n}A^{\SW}_{0}\right)\!\cdot\!\left(\nabla_{n}B^{\SW}_{0}\right)\right)(n) \right. \\ \nonumber
 &\hspace{1.5cm}\left. + i n\cdot\left(\left(\nabla_{n}A^{\SW}_{0}\right)\times\left(\nabla_{n}B^{\SW}_{0}\right)\right)(n)\right) \\ \nonumber
 &\hspace{0.5cm} + d^{-2}_{j}\left(A^{\SW}_{0}(n)B^{\SW}_{2}(n) + A^{\SW}_{1}(n)B^{\SW}_{1}(n) + A^{\SW}_{2}(n)B^{\SW}_{0}(n)\right. \\ \nonumber
&\hspace{1.5cm}\left. -\left(\left(\nabla_{n}A^{\SW}_{0}\right)\cdot\left(\nabla_{n}B^{\SW}_{1}\right)\right)(n)-\left(\left(\nabla_{n}A^{\SW}_{1}\right)\cdot\left(\nabla_{n}B^{\SW}_{0}\right)\right)(n) \right.\\ \nonumber
 &\hspace{1.5cm}\left. -3\left(\left(\nabla_{n}A^{\SW}_{0}\right)\cdot\left(\nabla_{n}B^{\SW}_{0}\right)\right)(n) \right. \\ \nonumber
 &\hspace{1.5cm}\left. + \tfrac{1}{2}\left(\left((n\times\nabla_{n})^{2}A^{\SW}_{0}\right)(n)B^{\SW}_{0}(n) + A^{\SW}_{0}(n)\left((n\times\nabla_{n})^{2}B^{\SW}_{0}\right)(n)\right) \right. \\ \nonumber
 &\hspace{1.5cm}\left. - \tfrac{1}{2}\left((n\times\nabla_{n})^{2}A^{\SW}_{0}\right)(n)\left((n\times\nabla_{n})^{2}B^{\SW}_{0}\right)(n) \right. \\ \nonumber
 &\hspace{1.5cm}\left. +\tfrac{1}{2}(n\times\nabla_{n})^{2}\left(\left(\nabla_{n}A^{\SW}_{0}\right)\cdot\left(\nabla_{n}B^{\SW}_{0}\right)\right)(n) \right. \\ \nonumber
 &\hspace{1.5cm}\left. -\tfrac{1}{2}\left(\left(\left(\nabla_{n}(n\times\nabla_{n})^{2}A^{\SW}_{0}\right)\!\cdot\!\left(\nabla_{n}B^{\SW}_{0}\right)\right)(n)\! +\! \left(\left(\nabla_{n}A^{\SW}_{0}\right)\!\cdot\!\left(\nabla_{n}(n\times\nabla_{n})^{2}B^{\SW}_{0}\right)\right)(n)\right)\right. \\ \nonumber
 &\hspace{1.5cm}\left. +i n\cdot\left(\left(\left(\nabla_{n}A^{\SW}_{0}\right)\times\left(\nabla_{n}B^{\SW}_{1}\right)\right)(n) + \left(\left(\nabla_{n}A^{\SW}_{1}\right)\times\left(\nabla_{n}B^{\SW}_{0}\right)\right)(n) \right. \right.\\ \nonumber
 &\hspace{1.5cm}\left.\left. - 6\left(\left(\nabla_{n}A^{\SW}_{0}\right)\times\left(\nabla_{n}B^{\SW}_{0}\right)\right)(n) \right.\right. \\ \nonumber
 &\hspace{1.5cm}\left.\left. + \tfrac{1}{2}\left(\left(\nabla_{n}(n\times\nabla_{n})^{2}A^{\SW}_{0}\right)\!\times\!\left(\nabla_{n}B^{\SW}_{0}\right)\right)(n)\! +\! \left(\left(\nabla_{n}A^{\SW}_{0}\right)\!\times\!\left(\nabla_{n}(n\times\nabla_{n})^{2}B^{\SW}_{0}\right)\right)\!(n) \right)\right. \\ \nonumber
 &\hspace{1.5cm}\left. -\tfrac{i}{2}(n\times\nabla_{n})^{2}\left(n\cdot\left(\left(\nabla_{n}A^{\SW}_{0}\right)\times\left(\nabla_{n}B^{\SW}_{0}\right)\right)(n)\right)\right) \\ \nonumber
 &\hspace{0.5cm} + \cO(d^{-3}_{j}).
\end{align}
\end{Remark}
Now, let us elaborate on the model \eqref{eq:spinorbitHamiltonian}:\\[0.1cm]
The symbol $H^{\lambda}_{d_{j}}$ of $\hat{H}^{\lambda}_{d_{j}}$ is\cite{VarillyTheMoyalRepresentation}:
\begin{align}
\label{eq:spinorbitHamiltoniansymbol}
H^{\lambda}_{d_{j}}(n) & = (1-\lambda)S_{3} + \lambda\sqrt{1-d^{-2}_{j}}n\cdot S \\ \nonumber
 & = \underbrace{(1-\lambda)S_{3}+\lambda n\cdot S}_{=H^{\lambda}_{0}(n)} + \lambda\sum_{k=1}^{\infty}\binom{2k}{k}(1-2k)^{-1}(4d_{j})^{-2k}n\cdot S.
\end{align}
Here, the semi-classical expansion is exact. Although, Faure and Zhilinskii use the lower symbol,
\begin{align}
\label{spinorbitlowersym}
L^{d_{j}}_{H^{\lambda}}(n)=(\zeta^{d_{j}}_{n},\hat{H}^{\lambda}_{d_{j}}\zeta^{d_{j}}_{n})_{\fH_{j}}=(1-\lambda)S_{3}+\lambda(1-d^{-1}_{j})n\cdot S,
\end{align}
w.r.t. a family of spin coherent states $\{\zeta^{d_{j}}_{n}\}_{n\in S^{2}}\subset\fH_{j}$, their analysis regards only the principal part $H^{\lambda}_{0}$, which is the same as ours. Therefore, all their findings apply to our case as well. Differences arise, when it comes of next-to-leading-order corrections, because the Stratonovich-Weyl symbol has contributions in all even orders of $d^{-1}_{j}$, while the lower symbol acquires only a first order contribution. In view of the previous subsection, this is especially interesting in the context of dynamics and the time-dependent Born-Oppenheimer approximation, because already the first order adiabatic theorem requires us to take the $d^{-1}_{j}$-order into account. A similar observation could be made, if we were to use the upper symbol.\\
Denoting the eigenvectors of $S_{3}$ by $\psi_{m},\ m=-s,...,s,$ we can state spectral properties of $H^{\lambda}_{0}$ in the following way:
\begin{align}
\label{eq:principalspectrumspin}
H^{\lambda}_{0}(n)\psi_{m}(n,\lambda) & = E_{m}(n,\lambda)\psi_{m}(n_{\lambda}), & \psi_{m}(n_{\lambda}) & = u^{\lambda}_{0}(n)^{*}\psi_{m}, \\ \nonumber
E_{m}(n,\lambda) & = N(n,\lambda)m, & (1-\lambda)e_{3}+\lambda n & = N(n,\lambda)n_{\lambda}, \\ \nonumber
N(n,\lambda) & = \sqrt{\lambda^{2}+(1-\lambda)^{2}+2\lambda(1-\lambda)\cos(\theta)}, & \cos(\theta) & =e_{3}\cdot n.
\end{align}
Thus, the spectrum of $H^{\lambda}_{0}(n)$ is non-degenerate for all $n\in S^{2},\ \lambda\in[0,1]$, with the exception $n=e_{3},\ \lambda = \frac{1}{2}$, where a collective degeneracy appears, $H^{\lambda=\frac{1}{2}}(-e_{3})=0$\cite{FaureTopologicalPropertiesOf}. Here, $u^{\lambda}_{0}(n)$ is a $SU(2)$-element corresponding to the rotation of $n_{\lambda}$ to $e_{3}$ via the adjoint action. It can be obtained explicitly, e.g. in ZYZ-notation (standard spherical coordinates relative to $\{e_{1},e_{2},e_{3}\}\subset\R^{3}$), as:
\begin{align}
\label{eq:zyzunitaryspin}
u^{\lambda}_{0}(n) & = e^{-i\varphi(n_{\lambda}) S_{3}}e^{i\theta(n_{\lambda})S_{2}}e^{i\varphi(n_{\lambda})S_{3}},
\end{align}
where the rotation angles can be read from $n_{\lambda}$ to be:
\begin{align}
\label{eq:zyzanglesspin}
\cos(\theta(n_{\lambda})) & = N(n,\lambda)^{-1}((1-\lambda)+\lambda\cos(\theta)), & \sin(\theta(n_{\lambda})) & = N(n,\lambda)^{-1}\sin(\theta), \\ \nonumber
\varphi(n_{\lambda}) & = \varphi(n_{\lambda=0}) = \varphi.
\end{align}
At this point, we should mention, that the spectral projections, $\pi^{\lambda}_{m,0}(n) = \psi_{m}(n_{\lambda})\otimes(\psi_{m}(n_{\lambda}),\ .\ )_{\fH_{s}}$, are smooth (in $n$) and globally defined for all $\lambda\in[0,\tfrac{1}{2})\cup(\tfrac{1}{2},1]$. Cleary, this is not the case for the unitary $u^{\lambda}_{0}(n)$, as can be deduced from the findings of Faure and Zhilinskii. Namely, every spectral projection $\pi^{\lambda}_{m,0}$ gives rise to a line bundle,
\begin{align}
\label{eq:adiabaticlinebundlespin}
\pi^{\lambda}_{m,0}\fH_{s} & \longrightarrow S^{2},
\end{align}
called the \textit{adiabatic line bundle of spectral index $m$}. The natural connection in each of the line bundles is given by the Berry-Simon connection,
\begin{align}
\label{eq:berrysimonspin}
A^{\lambda}_{m}(n) & = i(\psi_{m}(n_{\lambda}),d\psi_{m}(n_{\lambda}))_{\fH_{s}},
\end{align}
and its curvature,
\begin{align}
\label{eq:berrysimoncurvature}
F^{\lambda}_{m}(n) & = dA^{\lambda}_{m}(n),
\end{align}
gives the (first) Chern number of the line bundle,
\begin{align}
\label{eq:chernnumberspin}
C^{\lambda}_{m} & = \frac{1}{2\pi}\int_{S^{2}}F^{\lambda}_{m}\in\Z.
\end{align}
Its value was found by Faure and Zhilinskii to be:
\begin{align}
\label{eq:chernnumbervaluespin}
C^{\lambda}_{m} & = \left\{\begin{matrix} 0 & \lambda\in[0,\tfrac{1}{2}) \\ -2m & \lambda\in(\tfrac{1}{2},1] \end{matrix}\right. .
\end{align}
In this sense, the degeneracy at $\lambda=\frac{1}{2}$ is said to have a \textit{topological charge}\cite{FaureTopologicalPropertiesOf}. Since we are dealing with line bundles, the Chern number is a complete (topological) invariant\cite{BottDifferentialFormsIn}, i.e. only for $\lambda<\frac{1}{2}$ can the line bundles \eqref{eq:adiabaticlinebundlespin} be (smoothly) trivial for all $m=-s,...,s$\footnote{Vanishing of $C^{\lambda}_{m}$ only implies topological triviality of \eqref{eq:adiabaticlinebundlespin}, but one can explicitly show smooth triviality.}.\\
Thus, only for $\lambda<\frac{1}{2}$ can the unitary map $u^{\lambda}_{0}:S^{2}\rightarrow U(\fH_{s})$, be smooth and globally defined, and the program of space adiabatic perturbation theory can be made sense of ($u^{\lambda}_{0}$ is the natural candidate for the reference unitary\cite{PanatiSpaceAdiabaticPerturbation}). For $\lambda>\frac{1}{2}$, the program, presumably, has to be modified, e.g. by adapting the (de-)quantisation to the non-trivial Berry-Simon connection $A^{\lambda}_{m}$, as was done in \cite{FreundEffectiveHamiltoniansFor} for the case of magnetic, periodic Schr{\"o}dinger operators with non-trivial Bloch bands, i.e. non-trivial line bundles over the toric component in the Bloch-Floquet splitting of $\R^{d}$.\\
That the non-triviality of the line bundles \eqref{eq:adiabaticlinebundlespin} for $\lambda>\frac{1}{2}$, and the entailed non-existence of a globally smooth reference unitary $u^{\lambda}_{0}$, is not just a minor technical drawback, can be understood from the results of Faure and Zhilinskii, as well:\\[0.1cm]
They argue that the Chern number manifests itself in the exact spectrum of the Hamiltonian $\hat{H}^{\lambda}_{d_{j}}$ of the coupled system in the sense of a topological quantum number, which measures the dimension of the range of the projection, $\hat{\Pi}_{m}^{\lambda,d_{j}}$, onto the almost invariant subspace constructed from $\pi^{\lambda}_{m,0}$ (which still exists) in the limit $d_{j}\rightarrow\infty$:
\begin{align}
\label{eq:almostinvarianceandchern}
\dim\img\hat{\Pi}_{m}^{\lambda,d_{j}} & \underset{d_{j}\rightarrow\infty}{\sim}d_{j}-C^{\lambda}_{m} = (2j+1)+2m.
\end{align}
For $\lambda=1$, this formula gives the exact degeneracy of spectrum for (pure) spin-orbit coupling, i.e. the dimension of the dynamically stable subspaces.\\[0.1cm]
But, from the perspective of space adiabatic perturbation theory, the dimension of $\img\hat{\Pi}_{m}^{\lambda,d_{j}}$ would be forced to be:
\begin{align}
\label{eq:almostinvarianceandadibatic}
\dim\img\hat{\Pi}_{m}^{\lambda,d_{j}} & = d_{j} = 2j+1,
\end{align}
because of unitarily equivalence to the reference projection $\hat{\Pi}_{r} = \hat{U}^{\lambda,d_{j}}_{m}\hat{\Pi}^{\lambda,d_{j}}_{m}\left(\hat{U}^{\lambda,d_{j}}_{m}\right)^{*}$, which, by construction, satisfies:
\begin{align}
\label{eq:referencedimension}
\dim\img\hat{\Pi}_{r} & = d_{j}.
\end{align}
Therefore, if we were to apply space adiabatic perturbation theory to $\hat{H}^{\lambda}_{d_{j}}$ with $\lambda>\frac{1}{2}$, we would necessarily fail to predict the correct almost invariant subspaces for the dynamics. This is most prominently visible for (pure) spin-orbit coupling ($\lambda=1$).\\[0.25cm]
We conclude the discussion of the model by providing the first order expansion of the effective Hamiltonian symbol $h^{\lambda}_{m,(1)}$ restricted to the reference spaces in the case of $\lambda<\frac{1}{2}$ and $s=\frac{1}{2}$ (the fundamental representation of $SU(2)$:\\[0.1cm]
The Hamiltonian symbol is given in terms of the Pauli matrices:
\begin{align}
\label{eq:spinorbitHamiltonianPauli}
H^{\lambda}_{d_{j}}(n) & =  \frac{1}{2}\left((1-\lambda)\sigma_{3} + \lambda\sqrt{1-d^{-2}_{j}}n\cdot \sigma\right) \\ \nonumber
 & = \frac{1}{2}\left(\begin{matrix} (1-\lambda) + \lambda n_{3}\sqrt{1-d^{-2}_{j}} & \lambda (n_{1}-i n_{2})\sqrt{1-d^{-2}_{j}} \\ \lambda (n_{1}+i n_{2})\sqrt{1-d^{-2}_{j}} & -(1-\lambda) - \lambda n_{3}\sqrt{1-d^{-2}_{j}} \end{matrix}\right) \\ \nonumber
 & = \underbrace{\frac{1}{2}\left(\begin{matrix} (1-\lambda) + \lambda n_{3} & \lambda (n_{1}-i n_{2}) \\ \lambda (n_{1}+i n_{2}) & -(1-\lambda) - \lambda n_{3} \end{matrix}\right)}_{=H^{\lambda}_{0}(n)} + \cO(d^{-2}_{j}),
\end{align}
and its eigenvalues are
\begin{align}
\label{eq:spinorbiteigenvalues}
E^{\lambda}_{\pm}(n) & = \pm\frac{1}{2}N(n,\lambda),
\end{align}
which are globally separated by a gap $|E^{\lambda}_{+}(n)-E^{\lambda}_{-}(n)|\geq N(n,\lambda)\geq\min_{\theta\in[0,\pi)}N(n,\lambda)=:g^{\lambda}_{\frac{1}{2}}>0$, as long as $\lambda\neq\frac{1}{2}$ (fig. \ref{fig:spinorbitspectraldistancePauli}).
\\
The eigenvectors, corresponding to $m=\pm\frac{1}{2}$ (abbreviated: $\pm$), are (in standard spherical coordinates):
\begin{align}
\label{eq:spinorbiteigenvectorsPauli}
\psi_{+}(n_{\lambda}) & = \left(\begin{matrix} \cos(\tfrac{\theta(n_{\lambda})}{2}) \\ e^{i\varphi}\sin(\tfrac{\theta(n_{\lambda})}{2}) \end{matrix}\right), & \psi_{-}(n_{\lambda}) & = \left(\begin{matrix} -e^{-i\varphi}\sin(\tfrac{\theta(n_{\lambda})}{2}) \\ \cos(\tfrac{\theta(n_{\lambda})}{2}) \end{matrix}\right),
\end{align}
which are well-defined for $\lambda<\frac{1}{2}$.\\
For $\lambda>\frac{1}{2}$, these expression remain valid away from $\theta=\pi\ (n=-e_{3})$, where $\psi_{\pm}$ are not uniquely defined. The projections,
\begin{align}
\label{eq:spinorbiteigenprojectionsPauli}
\pi^{\lambda}_{+,0}(n) & = \left(\begin{matrix} \cos^{2}(\tfrac{\theta(n_{\lambda})}{2}) & e^{-i\varphi}\sin(\tfrac{\theta(n_{\lambda})}{2})\cos(\tfrac{\theta(n_{\lambda})}{2}) \\ e^{i\varphi}\sin(\tfrac{\theta(n_{\lambda})}{2})\cos(\tfrac{\theta(n_{\lambda})}{2}) & \sin^{2}(\tfrac{\theta(n_{\lambda})}{2}) \end{matrix}\right), \\ \nonumber
\pi^{\lambda}_{-,0}(n) & = \left(\begin{matrix} \sin^{2}(\tfrac{\theta(n_{\lambda})}{2}) & -e^{-i\varphi}\sin(\tfrac{\theta(n_{\lambda})}{2})\cos(\tfrac{\theta(n_{\lambda})}{2}) \\ -e^{i\varphi}\sin(\tfrac{\theta(n_{\lambda})}{2})\cos(\tfrac{\theta(n_{\lambda})}{2}) & \cos^{2}(\tfrac{\theta(n_{\lambda})}{2}) \end{matrix}\right),
\end{align}
are well-defined for all $\lambda\in[0,1]$ except $\lambda=\frac{1}{2}$, as explained above. The reference projections are provided by
\begin{align}
\label{eq:spinorbitreferenceprojectionsPauli}
\pi_{+,r} & = \pi^{\lambda=0}_{+,0} = \left(\begin{matrix} 1 & 0 \\ 0 & 0 \end{matrix}\right), & \pi_{-,r} & = \pi^{\lambda=0}_{-,0} = \left(\begin{matrix} 0 & 0 \\ 0 & 1 \end{matrix}\right),
\end{align}
and we choose
\begin{align}
\label{eq:spinorbitreferenceunitaryPauli}
u^{\lambda}_{0}(n) & = \left(\begin{matrix} \cos(\tfrac{\theta(n_{\lambda})}{2}) & e^{-i\varphi}\sin(\tfrac{\theta(n_{\lambda})}{2}) \\ -e^{i\varphi}\sin(\tfrac{\theta(n_{\lambda})}{2}) & \cos(\tfrac{\theta(n_{\lambda})}{2}) \end{matrix}\right)
\end{align}
as reference unitary. The Berry-Simon connection and curvature are:
\begin{align}
\label{eq:spinorbitBerrySimonPauli}
A^{\lambda}_{\pm}(n) & = \mp\frac{1}{2}\left(1-\cos(\theta(n_{\lambda}))\right)d\varphi, & F^{\lambda}_{\pm}(n) & = \mp\frac{1}{2}\sin(\theta(n_{\lambda}))d\theta(n_{\lambda})\wedge d\varphi.
\end{align}
In accordance with the general statements above, we find that $C^{\lambda<\frac{1}{2}}_{\pm}=0$ and $C^{\lambda>\frac{1}{2}}_{\pm}=\mp1$. Because the projections \eqref{eq:spinorbiteigenprojectionsPauli} correspond to single, non-degenerate eigenvalues, $E^{\lambda}_{\pm}(n) = \pm\frac{1}{2}N(n,\lambda)$, of $H^{\lambda}_{0}$, the effective Hamiltonian symbols are scalar, when restricted to the reference space, and the first order contributions, $h^{\lambda}_{\pm,1}$, can be found from the formula\cite{PanatiSpaceAdiabaticPerturbation}:
\begin{align}
\label{eq:firstorderHamiltonianPauli}
h^{\lambda}_{\pm,(1)}(n) & \sim h^{\lambda}_{\pm,0}(n) + d^{-1}_{j}h^{\lambda}_{\pm,1}(n) + \cO(d^{-2}_{j}),
\end{align}
\begin{align}
\label{eq:firstorderHamiltonianPauli0}
h^{\lambda}_{\pm,0}(n) & = E^{\lambda}_{\pm}(n)\pi_{\pm,r}, \\[0.25cm]
\label{eq:firstorderHamiltonianPauli1}
 h^{\lambda}_{\pm,1}(n) & = \pi_{\pm,r}((u^{\lambda}_{0}\star H^{\lambda}_{0})_{1}(n)-(E^{\lambda}_{\pm}\star u^{\lambda}_{0})_{1}(n))u^{\lambda}_{0}(n)^{*}\pi_{\pm,r} \\ \nonumber
 & = E^{\lambda}_{\pm}(n)\pi_{\pm,r}\left(\left((n\times\nabla_{n})^{2}u^{\lambda}_{0}\right)(n)u^{\lambda}_{0}(n)^{*}+u^{\lambda}_{0}(n)\left((n\times\nabla_{n})^{2}(u^{\lambda}_{0})^{*}\right)(n)\right)\pi_{\pm,r} \\ \nonumber
 &\hspace{0.5cm} + i\pi_{\pm,r}\left(n\cdot\left((\nabla_{n}u^{\lambda}_{0})\times(\nabla_{n}H^{\lambda}_{0})\right)(n)-n\cdot\left((\nabla_{n}E^{\lambda}_{\pm})\times(\nabla_{n}u^{\lambda}_{0})\right)(n)\right)u^{\lambda}_{0}(n)^{*}\pi_{\pm,r} \\ \nonumber
 & = E^{\lambda}_{\pm}(n)\pi_{\pm,r}\left((\nabla_{n}u^{\lambda}_{0})(n)\cdot(\nabla_{n}(u^{\lambda}_{0})^{*})(n)\right)\pi_{\pm,r} \\ \nonumber
 &\hspace{0.5cm} + \pi_{\pm,r}(2 i n)\cdot\left((\nabla_{n}E^{\lambda}_{\pm})\times(u^{\lambda}_{0}\nabla_{n}(u^{\lambda}_{0})^{*})\right)(n)\pi_{\pm,r} \\ \nonumber
 &\hspace{0.5cm} - \pi_{\pm,r}(i n)\cdot\left((\nabla_{n}u^{\lambda}_{0})(n)\times((H^{\lambda}_{0}-E^{\lambda}_{\pm})\nabla_{n}(u^{\lambda}_{0})^{*})(n)\right)\pi_{\pm,r} \\ \nonumber
 & = -E^{\lambda}_{\pm}(n)\pi_{\pm,r}\left((u^{\lambda}_{0}\nabla_{n}(u^{\lambda}_{0})^{*})(n)\!\cdot\!(u^{\lambda}_{0}\nabla_{n}(u^{\lambda}_{0})^{*})(n)\right)\pi_{\pm,r} \\ \nonumber
 &\hspace{0.5cm} + \pi_{\pm,r}(2 i n)\!\cdot\!\left((\nabla_{n}E^{\lambda}_{\pm})\times(u^{\lambda}_{0}\nabla_{n}(u^{\lambda}_{0})^{*})\right)\!(n)\pi_{\pm,r} \\ \nonumber
 &\hspace{0.5cm} + (E^{\lambda}_{\mp}(n)-E^{\lambda}_{\pm}(n))\pi_{\pm,r}(i n)\cdot\left((u^{\lambda}_{0}\nabla_{n}(u^{\lambda}_{0})^{*})(n)\times((1-\pi_{\pm,r})u^{\lambda}_{0}\nabla_{n}(u^{\lambda}_{0})^{*})(n)\right)\pi_{\pm,r} \\ \nonumber
 & =\bigg(\tfrac{1}{\sin(\theta)}\left(2(\partial_{\theta}E^{\lambda}_{\pm})(n)A^{\lambda}_{\pm}(n)_{\varphi}-(E^{\lambda}_{\mp}(n)-E^{\lambda}_{\pm}(n))F^{\lambda}_{\pm}(n)_{\theta\varphi}\right) \\ \nonumber
 &\hspace{0.5cm}+\tfrac{1}{\sin^{2}(\theta)}E^{\lambda}_{\pm}(n)\bigg(\left(\tfrac{2 E^{\lambda}_{\pm}(n)}{\lambda}F^{\lambda}_{\pm}(n)_{\theta\varphi}\right)^{2}\mp A^{\lambda}_{\pm}(n)_{\varphi}\bigg)\bigg)\pi_{\pm,r} \\ \nonumber
 & =\bigg(\tfrac{1}{\sin(\theta)}\left(2(\partial_{\theta}E^{\lambda}_{\pm})(n)A^{\lambda}_{\pm}(n)_{\varphi}+2E^{\lambda}_{\pm}(n)F^{\lambda}_{\pm}(n)_{\theta\varphi}\right) \\ \nonumber
 &\hspace{0.5cm} + \tfrac{1}{\sin^{2}(\theta)}E^{\lambda}_{\pm}(n)\bigg(\left(\tfrac{2 E^{\lambda}_{\pm}(n)}{\lambda}F^{\lambda}_{\pm}(n)_{\theta\varphi}\right)^{2}\mp A^{\lambda}_{\pm}(n)_{\varphi}\bigg)\bigg)\pi_{\pm,r}.
\end{align}
Here, we used in the first line of \eqref{eq:firstorderHamiltonianPauli1} that $H^{\lambda}_{1}=0$. Apart from this, the third line of \eqref{eq:firstorderHamiltonianPauli1} is still valid in general, and we observe that in addition to a familiar term containing the Berry-Simon connection (second term, cp. \eqref{eq:peierlsHamiltonianfirstorder}) two further terms appear, which can be attributed to the non-trivial geometry of $S^{2}$. The fourth to the sixth line are special to the model at hand, but we see that the Berry-Simon curvature already affects the first order contribution.
\section{Conclusions \& perspectives}
\label{sec:con}
In sections \ref{sec:boa} \& \ref{sec:weyl}, we have seen how the original Born-Oppenheimer ansatz, and its restricted applicability to slow-fast couplings via orthogonal pure state quantisations (fibred operators) in the analysis of multi-scale quantum systems, can be superseded by the more flexible space adiabatic perturbation theory, which is formulated by means of a suitable deformation quantisation, e.g. Weyl quantisation (for slow variables with a phase space isomorphic to $\R^{2d}$) or Stratonovich-Weyl quantisation (for spin systems, see section \ref{sec:spin}). Thus, the non-commutativity obstacle raised in \cite{GieselBornOppenheimerDecomposition} (see section \ref{sec:intro}) is lifted in way structurally enriching and conceptually refining the perturbative treatment of scale-separated, coupled quantum systems. It is worth to point out, that in space adiabatic perturbation theory the (classical) parameter space of the slow variables has the structure of a phase space, in contrast to the slow subsystem's configuration space appearing in the conventional Born-Oppenheimer approach to molecular quantum systems. In view of the possible extraction of quantum field theory on curved spacetimes from loop quantum gravity, the appearance of a phase space structure in the treatment of the slow/gravitational subsystem is advantageous, because a point in phase space corresponds to a spacetime metric via the effective classical time evolution arising in the semi-classical approximation of the coupled system. But, it is precisely a spacetime metric, which is necessary for the construction of a quantum field theory on a curved spacetime. Furthermore, this indicates that quantum field theory on curved spacetimes is expected to be of relevance to the fourth step, i.e. the semi-classical limit, of space adiabatic perturbation theory, when the latter is applied to loop quantum gravity, and not so much to the preceding three steps, which are dominated by kinematical considerations regarding the slow subsystem's phase space. To elaborate on the last statement, we notice that without invoking dynamics the correspondence between phase space point (initial data: spatial metric and extrinsic curvature) and time evolution trajectories (spacetime metric) is lost. Another interesting aspect of the phase space picture turning up in the semi-classical approximation, is, that the effective evolution equations are tied to a certain adiabatically decoupled subspace, which is constructed from a spectral subspace (in the fast subsystem) of the (partially) dequantised Hamiltonian. The upshot of it being, that, in applications to quantised gravity-matter systems, the above mentioned emergent spacetime metric depends on the choice of spectral subspace in fast sector. Such dependence of the spacetime metric on the spectral properties of the matter field is commonly referred to as a \textit{rainbow metric}\cite{LafranceGravitysRainbowLimits, MagueijoGravitysRainbow, AssanioussiRainbowMetricFrom}, and arises naturally in the context of space adiabatic perturbation theory.\\
The spin-orbit model discussed in section \ref{sec:spin} provides an idealised, though explicitly realised, testing ground for the solution of the non-commutativity problem, while simultaneously showing the interplay between non-trivial topological properties of the slow sector's phase space and the structure of the total Hamiltonian.\\
Establishing parts of the main mathematical toolbox necessary to implement a deformation \mbox{(de-)}quantisation and an associated symbolic calculus for loop quantum gravity and other models, that are based on projective limit phase spaces $\overline{\Gamma}=\varprojlim_{i\in I}\Gamma_{i}$  built from co-tangent bundles, $\Gamma_{i}=T^{*}G_{i},\ i\in I$, of compact Lie groups, $G_{i},\ i\in I$, is the main topic of our companion articles\cite{StottmeisterCoherentStatesQuantumII, StottmeisterCoherentStatesQuantumIII}.\\
In subsection \ref{subsec:csboa}, we discussed the possibility to employ a coherent state quantisation for the slow variables to generalise the original Born-Oppenheimer ansatz. Noteworthy, the use of lower symbols (partial traces w.r.t. coherent states projections) to obtain effective Hamiltonians in loop quantum gravity models was already put forward in \cite{SahlmannTowardsTheQFT1,SahlmannTowardsTheQFT2}, although a systematic way to connect information on the spectral problem of the effective Hamiltonians with the spectral analysis of the total Hamiltonian was not explored therein. As we have seen in section \ref{sec:weyl}, the existence of a $\star$-product is of vital importance to establish such a link, i.e. the (de-)quantisation of the slow sector has to be a (strict) deformation quantisation. Regarding the latter, we argue in our companion articles \cite{StottmeisterCoherentStatesQuantumII, StottmeisterCoherentStatesQuantumIII}, that coherent state quantisations are generically, i.e. in the case of a non-compact phase space for the slow variables, to singular to serve as a basis for deformation (de-)quantisation\cite{FollandHarmonicAnalysisIn}. This explains why we focus on the development of a less singular Weyl quantisation\cite{StottmeisterCoherentStatesQuantumII, StottmeisterCoherentStatesQuantumIII} as pointed out above.
Nevertheless, we devote a section of our second article\cite{StottmeisterCoherentStatesQuantumII} to the discussion of a new form of the Segal-Bargmann-Hall transform \cite{HallTheSegalBargmann}, because this unitary map, represented as a resolution of unity, is at the heart of a coherent state quantisation of the co-tangent bundle, $T^{*}G$, of a compact Lie group $G$, and thus fits into the general discussion of phase space quantisations and their relevance to generalised Born-Oppenheimer schemes (see subsection \ref{subsec:csboa}).
\begin{acknowledgments}
AS gratefully acknowledges financial support by the Ev. Studienwerk e.V.. This work was supported in parts by funds from the Friedrich-Alexander-University, in the context of its Emerging Field Initiative, to the Emerging Field Project ``Quantum Geometry’’.
\end{acknowledgments}
\bibliography{boa1.bbl, boa1Notes.bib}
\bibliographystyle{aipnum4-1}
\label{sec:ref}
\end{document}